\documentclass[twocolumn, astrosymb]{aastex631}

\usepackage{amssymb, amsmath}
\usepackage{CJKutf8}
\usepackage{bm}
\usepackage{hyperref}
\usepackage{fontawesome5}
\usepackage{academicons}
\usepackage{graphicx}
\usepackage{float}

\shorttitle{Outskirt Stellar Mass in Illustris-TNG Massive Galaxies}
\shortauthors{Xu et al.}

\graphicspath{{./}{figures/}}

\hypersetup{colorlinks=true,
            citecolor=MidnightBlue,
            linkcolor=MidnightBlue,
            filecolor=magenta,
            urlcolor=MidnightBlue}
\defcitealias{Huang2018b}{Paper~I}	
\defcitealias{Huang2018c}{Paper~II}	

\def\apjs{{ApJS}}

\def\photoi{\texttt{Photutils.isophote}}

\def\h{\hskip -3 mm}

\def\deg{$^{\circ}$}

\def\lax{{$\mathrel{\hbox{\rlap{\hbox{\lower4pt\hbox{${\sim}$}}}\hbox{$<$}}}$}}
\def\gax{{$\mathrel{\hbox{\rlap{\hbox{\lower4pt\hbox{${\sim}$}}}\hbox{$>$}}}$}}
\def\simlt{\lower.5ex\hbox{$\; \buildrel < \over {\sim} \;$}}
\def\simgt{\lower.5ex\hbox{$\; \buildrel > \over {\sim} \;$}}

\def\mstar{{$M_{\star}$}}
\def\logms{{$\log_{10} (M_{\star}/M_{\odot})$}}
\def\logmh{{$\log_{10} (M_{\mathrm{halo}}/M_{\odot})$}}

\def\logm10{{$\log (M_{\star,10\ \mathrm{kpc}}/M_{\odot})$}}
\def\logm30{{$\log (M_{\star,30\ \mathrm{kpc}}/M_{\odot})$}}
\def\logm50{{$\log (M_{\star,50\ \mathrm{kpc}}/M_{\odot})$}}
\def\logm100{{$\log (M_{\star,100\ \mathrm{kpc}}/M_{\odot})$}}

\def\mhalo{{$M_{\rm halo}$}}

\def\mh200b{{$M_{\mathrm{200b}}$}}
\def\mh200c{{$M_{\mathrm{200c}}$}}

\newcommand{\maper}[1]{\ensuremath{M_{\star, {#1}\ \rm kpc}}}

\newcommand{\mout}[2]{{$M_{\star, [#1, #2]}$}}

\newcommand{\maperre}[1]{\ensuremath{M_{\star, {#1} R_{\rm e}}}}
\newcommand{\moutre}[2]{{$M_{\star, [#1, #2] R_{\rm e}}$}}

\def\exsitu{{\textit{ex situ}}}
\def\ins{{\textit{in situ}}}
\def\exs{{\textit{ex situ}}}

\def\sigms{{$\sigma_{M_{\star}}$}}

\def\sigmh{{$\sigma_{\log M_{\mathrm{halo}}}$}}

\def\mdpl2{\texttt{MDPL2}}

\def\illustris{{\tt Illustris}}
\def\tng{{\tt IllustrisTNG}}

\def\topn{{Top-$N$}}
\def\arepo{{\tt AREPO}}

\def\dsigma{{$\Delta\Sigma$}}

\definecolor{LightGray}{gray}{0.85}
\definecolor{Tab1}{RGB}{114, 158, 206}
\definecolor{Tab2}{RGB}{255, 158,  74}
\definecolor{Tab3}{RGB}{103, 191,  92}
\definecolor{Tab4}{RGB}{174, 199, 232}
\definecolor{Tab5}{RGB}{255, 187, 120}
\definecolor{Tab6}{RGB}{152, 223, 138}
\definecolor{Tab7}{RGB}{255, 152, 150}
\definecolor{Tab8}{RGB}{197, 176, 213}
\definecolor{hpurple}{HTML}{7E16DF}

\newcommand{\code}[1]{\textbf{\texttt{#1}}} 
\newcommand{\change}[1]{\textcolor{black}{#1}} 
\newcommand{\changing}[1]{\textcolor{black}{#1}}

\begin{document}
\begin{CJK*}{UTF8}{gbsn}

\title{The Outskirt Stellar Mass of Low-Redshift Massive Galaxies is an Excellent Halo Mass Proxy in Illustris/IllustrisTNG Simulations}

\correspondingauthor{Song Huang}
\email{shuang@tsinghua.edu.cn, xus21@mails.tsinghua.edu.cn}

\author[0000-0002-4460-0409]{Shuo Xu (许朔)}
\affiliation{Department of Astronomy, Tsinghua University, Beijing 100084, China}

\author[0000-0003-1385-7591]{Song Huang (黄崧)}
\affiliation{Department of Astronomy, Tsinghua University, Beijing 100084, China}

\author[0000-0002-3677-3617]{Alexie Leauthaud}
\affiliation{Department of Astronomy and Astrophysics, UCO/Lick Observatory, University of California, 1156 High Street, Santa Cruz, CA 95064, USA}

\author[0000-0001-9568-7287]{Benedikt Diemer}
\affiliation{Department of Astronomy, University of Maryland, College Park, MD 20742, USA}

\author[0009-0003-8358-8320]{Katya Leidig}
\affiliation{Department of Astronomy, University of Maryland, College Park, MD 20742, USA}

\author[0000-0003-3843-7366]{Carlo Cannarozzo}
\affiliation{New York University Abu Dhabi, PO Box 129188, Abu Dhabi, United Arab Emirates}
\affiliation{Center for Astrophysics and Space Science (CASS), New York University Abu Dhabi}

\author[0000-0002-2897-6326]{Conghao Zhou (周丛浩)}
\affiliation{Physics Department, University of California, Santa Cruz, CA 95064, USA}
\affiliation{Santa Cruz Institute for Particle Physics, Santa Cruz, CA 95064, USA}

\begin{abstract}

    Recent observations suggest that the extended stellar halos of low-redshift massive galaxies are tightly connected to the assembly of their dark matter halos. In this paper, we use the \illustris{}, \tng{}100, and \tng{}300 simulations to compare how different stellar aperture masses trace halo mass. For massive central galaxies ($M_\star\geq 10^{11.2}M_\odot$), we find that a 2-D outskirt stellar mass measured between 50 to 100 kpc (\mout{50}{100}) consistently outperforms other aperture-based stellar masses. We further show that \mout{50}{100} correlates better with halo mass than the total amount of accreted stars (the \exs{} mass), which suggests that not all accreted stars connect to halo assembly equally. While the galaxy formation recipes are different between \illustris{} and \tng{}100, the two simulations yield consistent \exs{} outskirt fractions for massive galaxies ($\sim 70$\% in \mout{50}{100}). These results demonstrate the potential of using the outskirt stellar mass to deepen our understanding of \change{the} galaxy-halo connection in massive dark matter halos and trace dark matter halos better.

\end{abstract}


\section{Introduction} 
    \label{sec:intro}

    Low-redshift massive galaxies (e.g., $M_{\star} > 10^{11.2} M_{\odot}$ at $z < 0.5$) represent the most massive stellar systems ever created in our universe. 
    Although they typically show simple elliptical morphology (e.g., \citealt{Baldry2004ApJ}; \citealt{Vulcani2011}; \citealt{Buitrago2013MNRAS}) and are dominated by old and metal-rich stellar populations responsible for their global red colors (e.g., \citealt{Gallazzi2005MNRAS}; \citealt{Renzini2006}), these rare objects have experienced drastic structural transformation, star-formation quenching, and complex merging history (e.g., \citealt{Thomas2005ApJ}; \citealt{Behroozi2013ApJ}), which involves all the essential internal or environmental physical processes for galaxy formation and evolution.
    At the same time, according to the current cosmological models, galaxies formed in dark matter halos and evolved with them. 
    The mass, the clustering, and the complete assembly history of dark matter halos play fundamental roles in shaping the massive galaxies we observe today (e.g., \citealt{Kauffman1993MNRAS}; \citealt{Moster2018MNRAS}; \citealt{Behroozi2019MNRAS}; \citealt{Bose2019MNRAS}). 
    Therefore, an accurate galaxy-halo connection model is essential for a complete physical understanding of the formation of massive galaxies.
    Meanwhile, since massive galaxies tend to reside within massive dark matter halos, their low-$z$ counterpart is typically found in some of the universe's most prominent dark matter structures. These halos emerge from the rare high-density peaks of primordial density fluctuations (\citealt{Press1974ApJ}).

    \change{The abundance, dark matter density distribution, and clustering properties of these massive halos} are all essential cosmological probes that can shed light on the initial condition of the universe, the evolution of large-scale structure, and the nature of dark energy \& dark matter (e.g., \citealt{Evrard1989ApJ}; \citealt{Wang1998ApJ}; \citealt{Diemand2005MNRAS}; \citealt{Vikhlinin2009ApJ}; \citealt{Rozo2010ApJ}; \citealt{Abbott2020PRD}).
    \change{Therefore, the study of massive halos, based on the observations of massive galaxies, is becoming increasingly crucial in ambitious cosmological surveys since these galaxies serve as  the stellar ``tracers" of such halos (e.g. \citealt{Xhakaj2024MNRAS}; \citealt{Kwiecien2024})}.
    
    To take advantage of massive galaxies' potential in galaxy formation and cosmology, one must establish a halo mass ``proxy'' that can accurately and precisely predict halo mass based on single or multiple observed properties of galaxies or baryonic matter around galaxies. 
    \change{The necessity of using a proxy is that although weak gravitational lensing provides a direct approach to constrain the dark matter distributions}, it is not yet practical to estimate a single galaxy's dark matter halo mass (except for very nearby massive clusters' halos, e.g., \citealt{Hudson2015MNRAS}; \citealt{Mandelbaum2016MNRAS}). 
    Therefore, a well-performed halo mass proxy is essential for selecting massive halos, constraining models of the galaxy-halo connection, and even inferring cosmological parameters.

    The X-ray luminosity/temperature of the hot gas (e.g., \citealt{Reiprich2002ApJ}; \citealt{Voit2005ReMP}; \citealt{Vikhlinin2006ApJ}) and the strength of the Sunyaez-Zeldovich effect in submillimeter observations (e.g., \citealt{Bleem2015ApJS}) are reliable halo mass proxies for massive halos. 
    Nevertheless, it is more cost-effective to develop a halo mass proxy based on the properties of galaxies in these halos. 
    The number of galaxies above a certain luminosity/stellar mass threshold within a small physical scale (e.g., $\sim$ 1 Mpc) - or the richness - is often used as an excellent halo mass proxy, as it is closely related to the subhalo abundance (e.g., \citealt{Andreon2010MNRAS}; \citealt{Murata2018ApJ}). 
    The richness of the quenched galaxy population in massive halos is often preferred and is typically obtained from optical data. This population forms a "red sequence" in the magnitude-color space, making the galaxies easier to identify.
    \change{This ``red-sequence'' richness method, used in algorithms such as {\tt redMaPPer} (e.g., \citealt{Rykoff2014ApJ}; \citealt{Rozo2014ApJ}) and {\tt CAMIRA} (e.g., \citealt{Oguri2014MNRAS})}, is now considered one of the best ways to select and measure the optical manifestation of low-$z$ massive halos - galaxy clusters and massive groups - in modern imaging surveys and has become the foundation of recent cluster cosmology constraints (e.g., \citealt{Costanzi2021PhRvD}). 
    Unfortunately, richness-based proxies also have serious systematics, such as projection bias, meaning that the projection of large-scale structures around the massive halo or overlapping multiple halos along the line of sight can significantly bias richness measurements (e.g., \citealt{Zu2017MNRAS}; \citealt{Costanzi2019}; \citealt{Wu2022MNRAS}).
    Therefore, complementary approaches are still much needed. 
    In recent years, the total luminosity/stellar mass of galaxies in clusters above a magnitude limit (e.g., \citealt{Yang2007ApJ}; \citealt{Palmese2020MNRAS}; \citealt{Tinker2021MNRAS}), the line-of-sight velocity dispersion of galaxies (e.g., \citealt{Serra2011MNRAS}; \citealt{Farahi2016MNRAS}), and the stellar velocity dispersion of the central galaxy (e.g., \citealt{Zahid2018ApJ}) have all been proposed as potential halo mass proxies.
    However, these candidates either suffer from projection bias or require more ``expensive'' spectroscopic observations.

    Imaging data of massive halos often reveal a dominant central galaxy, which is usually identified as the brightest cluster/group galaxy (BCG/BGG; e.g., \citealt{Dubinski1998ApJ}; \citealt{Laine2003AJ}). 
    \change{Theoretical models and simulations suggest that the growth of the central galaxy is closely related to the assembly history of its host halo (e.g., \citealt{DeLucia2007MNRAS}; \citealt{Behroozi2019MNRAS}).} 
    This physical connection gives rise to the well-known stellar-halo mass relation (SHMR), which indicates that the stellar mass of a galaxy statistically correlates with its dark matter halo mass (e.g., \citealt{Tinker2017ApJ}; see \citealt{Wechsler2018ARA&A} for a recent review).
    The SHMR of massive central galaxies can be defined straightforwardly and has been the focus of investigations into their galaxy-halo connection model (e.g., \citealt{Kravtsov2018, Erfanianfar2019, Golden-Marx2022}).    
    Conventionally, the stellar mass of a massive galaxy is measured within a specified aperture that encompasses the central regions of galaxies or using a simplified two-dimensional model. 
    \change{Despite that the \emph{average} SHMR of massive central galaxies can be observationally constrained, the scatter of halo mass at a fixed stellar mass is too large to make \changing{the} SHMR a useful halo mass indicator (e.g., see Fig.~5 in \citealt{Wechsler2018ARA&A}).} 
    A recent study by \citet{Huang2022} uses deep images of $0.2<z<0.5$ massive galaxies \change{($M_{\star} > 10^{11.2} M_{\odot}$)} from the Hyper Suprime-Cam (HSC) survey and their stacked galaxy-galaxy lensing signals. \changing{The $M_{\star} > 10^{11.2} M_{\odot}$ threshold ensures that we can study their extended stellar mass distributions within the selected redshift range. At the same time, above $10^{11.2} M_{\odot}$, the vast majority of the sample are elliptical galaxies that are also centrals.} It found that the performance of a stellar mass measurement as a halo mass proxy depends on the physical scale it covers, \changing{which is similar to the results in recent studies (e.g. \citealt{Moster2018MNRAS}, \citealt{Golden-Marx2019ApJ}, \citealt{Golden-Marx2023MNRAS}, \citealt{Golden-Marx2025MNRAS}).}
    The stellar mass estimated within a large aperture (e.g., 100 kpc) is a much better halo mass proxy than the one based on a smaller aperture (e.g., 10 kpc). 
    Surprisingly, the stellar mass at the outskirts of massive galaxies (e.g., between 50 and 100 kpc) is an even better halo mass proxy than the stellar mass estimated using a larger aperture (e.g., 100 kpc). 
    This result may be due to the ``two-phase'' formation of these massive galaxies (e.g., \citealt{Naab2009ApJl}; \citealt{Oser2010ApJ}; \citealt{Johansson2012ApJ}; \citealt{Hilz2013MNRAS}), where the majority of the stars formed outside of the main progenitor's halo and were later accreted into the system during mergers with satellite galaxies. 
    \changing{
    This \exs{} component, which is defined as the stars formed outside the main progenitor's halo and then accreted into the galaxies (e.g., \citealt{RodriguezGomez2016MNRAS}), dominates the stellar halos of massive galaxies and correlates better with the halo mass (e.g., \citealt{Bradshaw2020MNRAS})}. 
    The outskirt stellar mass of the central galaxy should be much less affected by projection bias. 
    If it is a comparable halo mass proxy to the richness, it can be used as an alternative way to identify cluster halos and measure their mass (e.g., \citealt{Xhakaj2024MNRAS}; \citealt{Kwiecien2024}).

    \change{On the other hand, the outskirts of massive galaxies are often considered part of the intra-cluster light (ICL) or intra-group light (IGrL) (e.g., \citealt{Montes2022NatAs}; \citealt{Contini2021Galax}) in the literature. These diffuse stellar components primarily originate from stars that were either tidally stripped or completely disrupted from satellite galaxies (e.g., \citealt{Murante2007MNRAS}; \citealt{Contini2014MNRAS}) and have been shown to correlate with the properties of their host halos (e.g., \citealt{Montes2019MNRAS}; \citealt{Sampaio-Santos2021MNRAS}; \citealt{Golden-Marx2023MNRAS}).}

    Therefore, in this work, \change{we use massive galaxies from the state-of-the-art hydrodynamical galaxy formation simulations}, \illustris{} and \tng{}, of galaxy formation with different physical recipes and resolutions to compare with the results from \citet{Huang2022}. 
    We will systematically evaluate the performance of different stellar masses as potential halo mass proxies. 
    Moreover, we will use the physically separated \ins{} and \exs{} components to investigate the physical origin of the outskirt stellar mass' better performance as a halo mass proxy. 

    This paper is organized as follows.
    Section \ref{sec:data} introduces the hydrodynamic simulations used in this work. 
    Section \ref{sec:methods} describes the technique for extracting stellar mass profiles of the simulated massive galaxies, fitting their SHMR. 
    We present our main results in Section \ref{sec:results} and provide detailed discussions of the physical implications and potential caveats in Section \ref{sec:discussions}. 
    Finally, Section \ref{sec:summary} summarizes the main conclusions of this work.

    The \illustris{} project adopted the $\Lambda$CDM cosmology with parameters $\Omega_m=0.2726$,~$\Omega_{\Lambda}=0.7274$,~$\Omega_b=0.0456$,~$H_0=70.4$~km~s$^{-1}$ Mpc$^{-1}$, 
    while the \tng{} project adopted different parameters $\Omega_m=0.3089$,~$\Omega_{\Lambda}=0.6911$,~$\Omega_b=0.0486$,~$H_0=67.74$~km~s$^{-1}$ Mpc$^{-1}$. 
    These two simulations adopted a \citet{Chabrier2003PASP} initial mass function.
    For dark matter halo mass, we adopt $M_{200c}$ -- the overdensity mass within a sphere where the average density is 200 times the critical density of the universe for all simulations.


\section{Simulation Data} 
    \label{sec:data}
    
    In this section, we briefly introduce the hydrodynamical simulations used in this work: the \illustris{}-1 simulation from the \illustris{} Project\footnote{\url{https://www.illustris-project.org/}} and the {\tt TNG100}, {\tt TNG300} simulations from the \tng{} Project\footnote{\url{https://www.tng-project.org/}}.
    
\subsection{The \illustris{} and \tng{} Simulations} 
    \label{ssec:illustrisandTNG}

    The \illustris{} project is a large cosmological simulation of galaxy formation using the moving mesh algorithm \arepo{} (\citealt{Springel2010MNRAS}, \citealt{Pakmor2011MNRAS}, \citealt{Weinberger2020ApJS}).
    It implements sophisticated physical recipes about gas cooling and photo-ionization, \change{star formation \& ISM, stellar evolution \& feedback, and supermassive black hole (SMBH) feedback} \citep{Vogelsberger2014MNRAS}.
    \change{The} \illustris{} simulation has helped explore a wide range of topics related to galaxy formation and evolution, including the assembly of massive galaxies (e.g., \citealt{Pillepich2014MNRAS}, \citealt{Cook2016ApJ}).
    Meanwhile, there are also well-known and substantial disagreements between \illustris{} and \change{observations} (\citealt{Genel2014MNRAS}; \citealt{Sparre2015MNRAS}; \citealt{RodriguezGomez2015MNRAS}) caused by incomplete or imperfect physical recipes.

    As the successor to the \illustris{} project, the \tng{} project is a suite of large-scale cosmological hydrodynamical simulations of galaxy formation using \emph{The Next Generation} of recipes for different physical processes (e.g., \citealt{Pillepich2018aMNRAS}).
    Compared to the \emph{original} \illustris{} series, \tng{} employs a magneto-hydrodynamical (MHD) simulation framework, adds a new kinetic SMBH feedback mode, and improves the treatment of the galactic wind, among many other changes and improvements (e.g., \citealt{Weinberger2017MNRAS}).
    These updates, especially the kinetic feedback mode, which operates at a low accretion rate, are crucial for following the evolution of massive galaxies and producing a more realistic population of galaxies relevant to this work (see Fig.~14 in \citealt{Pillepich2018MNRAS}). 

    \illustris{} includes a series of simulations with different resolutions and the same volume of $106.5^3$ Mpc$^3$.
    The \tng{} simulations come in three simulation volumes: {\tt TNG50} (with a box size of 51.7 comoving Mpc), {\tt TNG100} (110.7 cMpc) and {\tt TNG300} (302.6 cMpc).
    Each volume has a series of simulations with different mass resolutions for dark matter and baryons. 

    Although the \tng{} project has superseded \illustris{} in many ways, we chose to include both \illustris{} and \tng{} in this work to investigate the robustness of the performance of halo mass proxies against different underlying physical recipes. We adopt the highest resolution version of the \illustris{}, {\tt TNG100}, and {\tt TNG300} simulations. 
    \illustris{}-1 simulation has a dark matter mass resolution of $m_{\rm DM}= 6.3\times10^6~M_\odot$, an initial baryonic mass resolution of $m_{\rm baryon} = 1.3\times10^6~M_\odot$, and a gravitational softening scale of 710 pc at $z=0$, while {\tt TNG100-1} ({\tt TNG300-1}) has a dark matter mass resolution of $7.5\times10^6~M_\odot$ ($5.9\times10^7~M_\odot$), \change{a baryonic particle} resolution of $1.4\times10^6~M_\odot$ ($1.1\times10^7~M_\odot$), and a softening length of 740 pc (1480 pc).

    {\tt TNG100-1} shares the same initial conditions as the original simulation \illustris{}, making it perfect for comparing the impact of different physical recipes.
    {\tt TNG300-1} represents the largest volume from the \tng{} series, which can sample the high-mass end of halo and stellar mass functions better than {\tt TNG100-1} and provide a statistically significant sample of massive galaxies to describe \change{the SHMR} and evaluate different halo mass proxies.
    At the same time, \change{the physical resolution of the {\tt TNG300-1}} is still sufficient to characterize the detailed stellar mass distributions of massive galaxies at the physical scale relevant to \change{a comparison with HSC or other ground-based imaging data (e.g., at $>$ 10 kpc).}
    Comparison between {\tt TNG100-1} and {\tt TNG300-1} can also help investigate the resolution's impact on halo mass proxies' performance. 
    The physical resolution of the simulation can impact the galaxy-halo connection and stellar properties of massive galaxies in many subtle ways. 
    We refer the reader to Leidig et al. (in preparation) for a more comprehensive analysis. 

    Using the $\log_{10}[M_{\star}/M_\odot]\geq11.2$ total stellar mass\footnote{\change{The stellar mass here is the mass of all the stellar \changing{particles} in the subhalo in simulation}} cut, we selected 339 massive central galaxies from \illustris{}, 235 from {\tt TNG100-1}\footnote{\changing{Since \illustris{} and \tng{} use different galaxy formation recipes and the star formation in \illustris{} is more efficient, there is more massive galaxies in \illustris{} for the same stellar mass threshold and similar volume.}} and 2713 from {\tt TNG300-1} \change{at $z=0.4$, which corresponds to the median redshift of the HSC massive galaxy sample}. 
    In both \illustris{} and \tng{}, a central galaxy is defined as the most bound subhalo in a larger friends-of-friends (FOF) group.
    Although this definition is different from the large aperture stellar mass adopted by the massive galaxy sample of HSC, in \citet{Ardila2021MNRAS}, the authors showed that the stellar mass functions between observation and \illustris{} (and also \tng{}) are similar enough for a meaningful comparison.
    We excluded satellite galaxies to make it easier to characterize the SHMR of massive galaxies. 

\subsection{Stellar Mass Maps} 
    \label{ssec:maps}

    Generally, massive galaxies are 3-D objects with diverse intrinsic shapes and mass distributions.
    Therefore, it is natural that many previous works about low-$z$ massive galaxies using hydrosimulations choose to characterize the stellar mass distributions using 3-D spherical shells (e.g., \citealt{RodriguezGomez2016MNRAS} using \illustris{}; \citealt{Pillepich2018MNRAS} using \tng{}). 
    However, a perfect spherical shell only provides a biased view of the 3-D mass distribution. 
    More importantly, it is impossible to compare these 3-D profiles with actual observations, given the difficulty in inferring the intrinsic shape of massive galaxies (e.g., \citealt{Mendez-Abreu2016ASSL}; \citealt{Li2018ApJ}; \citealt{Bassetts2019MNRAS}).

    Following the logic and methods of \citet{Ardila2021MNRAS} and \citet{Cannarozzo2023MNRAS}, this work adopts the projected 2-D stellar mass maps to facilitate a comparison with the $M_{\star}$-based halo mass proxies estimated on HSC images. 
    Using the same strategy, for each massive galaxy in the \illustris{}, {\tt TNG100-1}, and {\tt TNG300-1} simulations, we project the positions of their stellar particles to a $300\times300$-pixels map along the three primary axes of the simulation boxes ({\tt XY}, {\tt YZ} and {\tt XZ}) regardless of the intrinsic shape or orientation of the galaxy. 
    \change{With a $1{\rm~kpc}$ pixel size, each map corresponds to a $300{\rm~kpc}$ box region around a massive galaxy and is large enough to capture all the stellar content that is relevant to current imaging observations for individual massive galaxies in HSC (\citealt{Huang2018})}.
    The maps in different projections can help us qualitatively evaluate the galaxy shape \& orientation's impact on each halo mass proxy's performance. 
    Since we only consider massive central galaxies, the stellar mass map includes all stellar particles from the Friend-of-Friend (FoF) group after removing the stellar particles from the satellite galaxies.
    And, using the definitions given by \citet{RodriguezGomez2016MNRAS}, we isolate the \ins{} and \exs{} stars in massive galaxies and create separated maps for them to help us investigate the underlying causes of different stellar masses' performance as halo mass proxies.

    In total, for each simulated massive galaxy, we create nine stellar mass maps: the maps for the total, \ins{}, and \exs{} stellar mass in the {\tt XY}, {\tt YZ} and {\tt XZ} projections. 
    We refer the authors to \citet{Ardila2021MNRAS} for more details of these stellar maps. 

    These maps are \emph{not} mock images with realistic imaging systematics and noise. 
    We adopted this strategy to focus on the \emph{intrinsic} behaviors of different stellar masses without worrying about the complications of real images. 
    Recently, \citet{Bottrell2024MNRAS} released realistic mock HSC images of \tng{} galaxies. 
    Unfortunately, it does not include the {\rm TNG300-1} galaxies, the primary data source for this work. We leave a detailed discussion on the impact of observational effects for future work, where we will further analyze mock-observed stellar halos in IllustrisTNG (Leidig et al., in prep).


\section{Methods} 
    \label{sec:methods}
    
    This section briefly describes the methods for measuring different stellar masses.
    First, we performed a 1-D isophotal analysis of the stellar mass maps to estimate the average ellipticity and position angle to define an elliptical aperture that describes the average isophotal shape of a massive galaxy.
    Then, we estimate several characteristic radii, such as the effective radius ($R_{\rm e}$), the radius enclosing 50\% of the total stellar mass, based on the stellar mass distribution.
    Finally, we define a series of elliptical apertures based on the values of physical (e.g., 10 or 100 kpc) or characteristic radii (e.g., $2 \times R_{\rm e}$) and measure the stellar masses within or between these apertures using the average isophotal shape. 
    We apply the same methods to the three simulations used in this work.

    \begin{figure*}[htb]
    \centering
    \includegraphics[width=1.05\linewidth,trim=6cm 0 0 0]{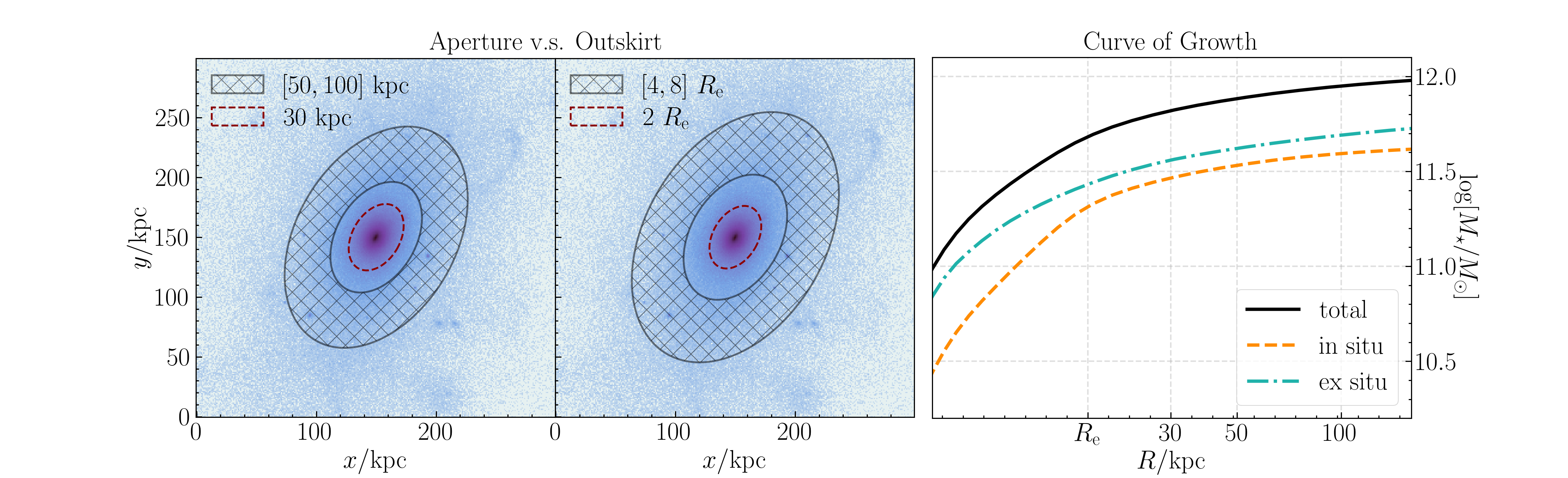}
    \caption{
        Definitions of aperture and outskirt stellar masses. 
        {\bf The left panel} shows \change{the 2-D stellar mass map of galaxy ID $=31188$ ($M_{\rm halo}=10^{14}~M_\odot$)} in the {\tt TNG100} simulation in the $\rm XY$ projection.
        The open inner ellipse (red, dashed line) defines the 30 kpc aperture for $M_{\star, 30}$.
        The outer annulus (black, hatched) defines the outskirt region used for $M_{\star, [50,100]}$.  
        {\bf Middle panel}: similar to the left panel, but now we define the aperture and outskirt regions using $R_{\rm e}$ ($\sim$ 14 kpc for this galaxy). The inner ellipse defines the region for $M_{\star, 2~R_{\rm e}}$ and the outer hatched annulus defines the region for $M_{\star, [4, 8]~R_{\rm e}}$.
        {\bf Right panel}: the curve of growth (CoG) for the total stellar mass distribution of this galaxy (solid black line) along with the CoGs of the \ins{} (orange, dashed line) and \exs{} (cyan, dot-dashed line) components for forced measurement.
        The \texttt{Jupyter} notebook for reproducing this figure can be found here:  \href{https://github.com/Xuchuyi/HaloMassProxy/blob/main/FigureNotebook/Fig1.ipynb}{\faGithub}.
        }
    \label{fig:fig1}
    \end{figure*}

\subsection{Isophotal Fitting and Curve of Growth}
    \label{ssec:profile}

    To define an appropriate elliptical isophotal shape for measuring stellar masses, we first perform a 1-D isophotal analysis on the stellar mass maps to convert the 2-D mass distribution into 1-D profiles of surface stellar mass density, ellipticity, and position angles.
    Above $10^{11.2} M_{\odot}$, the observed and simulated massive galaxies always exhibit simple morphology and smooth stellar mass distributions that a series of elliptical isophotes can adequately approximate. 
    We adopt the methodology laid out in \citet{Huang2018} and \citet{Li2022} for HSC images, which is also similar to the one used in \citet{Ardila2021MNRAS} and \citet{Cannarozzo2023MNRAS} for \tng{} galaxies. 
    As in \citet{Li2022}, we switch from the \texttt{IRAF} \texttt{Ellipse} procedure to its \texttt{Python} equivalent -- the {\tt isophote} module in \photoi{}, whose underlying algorithms are almost identical.

    We run {\tt isophote} with a step size of $=0.2$ in relative flux units after fixing the galaxy's centroid position at the map's center. 
    Since we removed satellite galaxies and other contaminations when generating the 2-D maps, we did not apply the object mask during the {\tt isophote} fitting. We took the mean stellar mass density value along each isophote. 
    As the first step, we allowed the geometry of the isophote to vary to derive the 1-D ellipticity and position angle profiles.
    We then calculated the intensity-weighted mean ellipticity and position angle within a galaxy-dependent radial range to define the average isophotal shape. 
    The inner boundary of the radial range is 8 kpc for the {\tt TNG300}\footnote{For 123 galaxies in {\tt TNG300} sample, because some stellar maps of these galaxies have little stellar particles or some asymmetric structure in the outer region, or their outer boundaries are smaller than 8 kpc, we change their inner boundary to 5 kpc}, and 3 kpc for {\tt TNG100} and \illustris{}.
    Regarding the outer boundary, we find that beyond a certain radius, the isophote shapes remain fixed at their initial values, indicating unreliable fitting at these distances. The reason is that the sparse distribution of stellar particles and asymmetric tidal structures make a meaningful elliptical isophotal fitting challenging \change{beyond} this boundary. Therefore, the outer boundary is defined at the radius where the isophotal fitting becomes unreliable. 
    In \citet{Huang2018}, the authors extracted the 1-D surface brightness profile along the semimajor axis of HSC massive galaxies using the average isophotal shape before integrating it to calculate the aperture and outskirt stellar mass. 
    \change{On simulated 2-D stellar mass maps, we instead directly perform elliptical aperture photometry using {\tt photutils} based on the average isophotal shape at different radii to form a ``curve of growth" (CoG), as illustrated in the right panel of Fig. \ref{fig:fig1}}, which describes the increasing trend of stellar mass enclosed in the elliptical apertures at a larger and larger radius. From a CoG, we can easily calculate different aperture and outskirt stellar masses. Assuming that we successfully exclude satellite galaxies from the stellar mass map, this direct approach provides a more accurate estimate of the aperture or outskirt stellar masses.
    Although this aperture photometry method is often not practical in real images, it helps to evaluate the intrinsic behavior of different stellar mass measurements as halo mass proxies in \change{simulations}. Furthermore, \change{these two methods provide consistent results when there is no contamination.}

    We also apply the same average isophotal shape derived using the total stellar mass distributions to measure the CoGs of the \ins{} and \exs{} components. \change{The measurements of these two distinct components can help us understand the physical mechanism that drives the performance of different \mhalo{} proxies.} Here, we ignore the differences between the average isophotal shapes of the \ins{} and \exs{} components. We will address such differences in future work, as they may reveal insight into the assembly history of massive galaxies.

    Meanwhile, as the intrinsic 3-D shape of a massive galaxy is rarely close to spherical, the direction of the 2-D projection leads to a variation of the 2-D shape, affecting the aperture/outskirt mass measurements. To explore this issue, we also perform the above procedures independently on all three projections of each galaxy ({\tt XY}, {\tt YZ} and {\tt XZ}). We briefly discuss this projection effect in \S\ref{ssec:proj}. 

    Moreover, suppose that we define the ``total stellar mass" as the sum of all the stellar particles on the 2-D map. In that case, we can estimate a series of characteristic radii based on the fraction of total stellar mass using CoG. For example, we define a galaxy's effective radius ($R_{\rm e}$) as the semimajor axis of the elliptical aperture enclosing $50$ percent of the total stellar mass.
    
    In Figure \ref{fig:fig1}, we demonstrate the average isophotal shape estimated for galaxy ${\rm ID}=31188$ in {\tt TNG100}, along with the CoGs of its total, \ins{}, and \exs{} components on the {\tt XY} projection map.
  
\subsection{Aperture and Outskirt \texorpdfstring{\mstar{}}{Mstar}}
    \label{ssec:cog}

    We measure two stellar mass types based on the 1-D CoGs described earlier. The \emph{aperture mass} is the stellar mass enclosed by an elliptical aperture defined by a characteristic semimajor axis length and an average isophotal shape. The \emph{outskirt mass} describes the stellar mass between two elliptical apertures with the same shape (i.e., the stellar mass difference between two aperture stellar masses). We want to emphasize that these definitions are not motivated by any ``physical'' meaning but because they are straightforward to implement in imaging data without model assumptions that can be sensitive to data quality and other systematic issues.
        
    As in \citet{Huang2022}, we can define an aperture or outskirt stellar mass in units of kpc (absolute physical size) or $R_{\rm e}$ (relative to the effective radius). The physical unit provides an unambiguous definition that is straightforward to implement on real data. However, the choice of physical size is subjective, if not arbitrary (e.g., 10 vs. 15 kpc). More importantly, the physical meaning of an aperture or outskirt mass using a physical unit could vary significantly with the galaxy's total stellar mass, affecting our understanding of their connection to the assembly history. However, apertures and outskirts defined in $R_{\rm e}$ have the advantage of scaling naturally with the intrinsic stellar mass distribution. On the other hand, the estimation of $R_{\rm e}$ in real images often depends on the image quality and, in particular, on the choice of photometric methods, models, and filters used in the observations. In light of this trade-off, we therefore decided to evaluate both approaches. 

    \change{For aperture stellar masses using physical units, we define a series of apertures at 10, 30, 50, 75, 100, and 150 kpc.} For example, \maper{30} stands for \mstar{} within a 30 kpc aperture. For radii in relative units, we define another series of apertures at 3, 4, 5, and $8\times R_{\rm e}$ (e.g., \maperre{4} refers to \mstar{} enclosed within four times the effective radius). We then use the difference between the two aperture stellar masses in the same series to define the corresponding outskirt mass (e.g., \mout{50}{100} means \mstar{} between the 50 and 100 kpc apertures; \moutre{3}{5} denotes \mstar{} between three and five times the effective radius). Motivated by \citet{Huang2022}, we focus on relatively large apertures to probe the SHMR of massive galaxies. We use 10 kpc (30 kpc) as the smallest aperture size for the \illustris{} and {\tt TNG100} ({\tt TNG300}) simulations. These choices help to prevent the \mstar{} measurements from being influenced by the force resolutions of the simulations. For the \mstar{} defined in $R_{\rm e}$, we only discuss the aperture or outskirt \mstar{} defined at $\geq 3\times R_{\rm e}$. \change{This choice is based on the median effective radius $R_{\rm e}$ ($\sim13{~\rm kpc}$)} of the {\tt TNG300} sample. Such a lower limit helps to ensure that most of the measurements remain unaffected by resolution constraints. Although the data differ, we choose many of the same aperture definitions used in \citealt{Huang2022} to foster a (semi)qualitative comparison. We illustrate the definitions of aperture and outskirt stellar masses in the left panels of Figure \ref{fig:fig1}. 
    
    As mentioned earlier, we measure the aperture and outskirt masses in all three projections for the same galaxy. We independently evaluate their performance as halo mass proxies, and we discuss the impact of projection in \S\ref{ssec:proj}. We also measure these masses separately for the \ins{} and \exs{} components. Note that we ignore the radial variations of the isophotal shape or the difference in shape among the different components. This choice is again motivated by typical photometric procedures in real data. We will briefly discuss its implications in \S\ref{ssec:exsscat}. The stellar mass distributions of most massive galaxies discussed here extend beyond 150 kpc. To account for the stellar content in the extremely low-density regime and investigate its relation to the dark matter halo requires a more careful treatment of the stars in satellite galaxies and several systematics in the simulation (e.g., the definition of halo boundary and mass resolution of the simulation). \change{A number of works have studied this regime (e.g., \citealt{Zhang2019ApJ}; \citealt{Kluge2024}; \citealt{Brown2024MNRAS}) and we refer the reader to Leidig et al. (in preparation) for further discussion.}

    \begin{figure*}[htb]
        \centering
        \includegraphics[width=1.05\linewidth,trim=3cm 0 0 0]{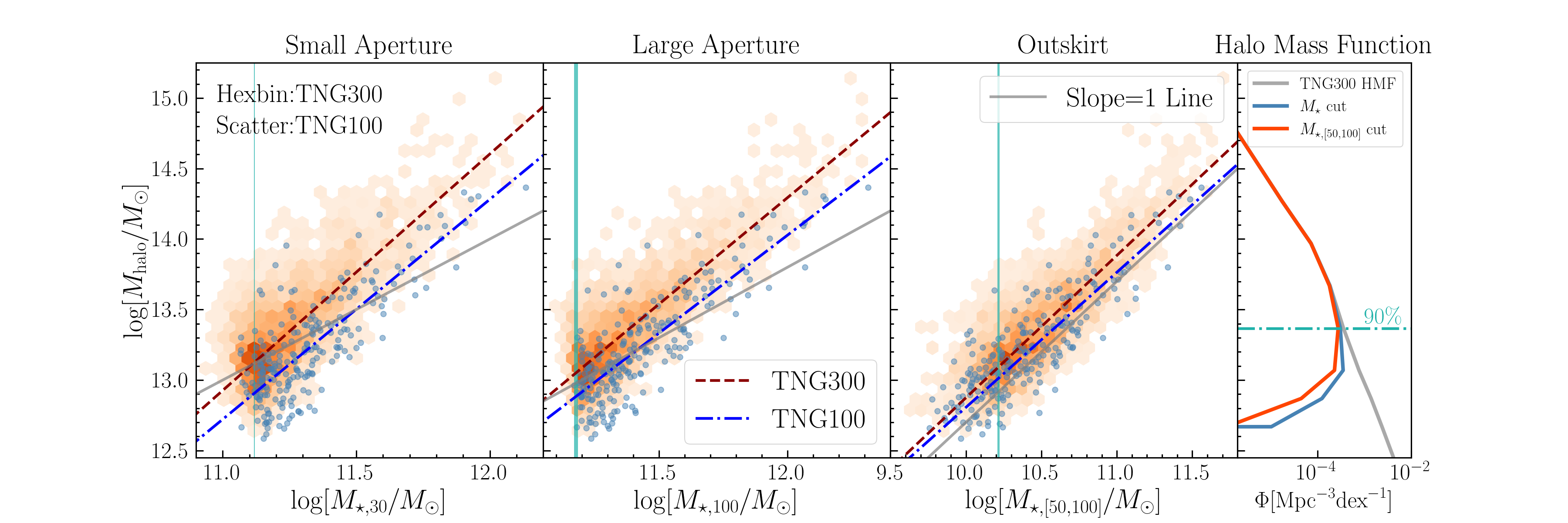}
        \caption{
            The distributions of massive galaxies from the {\tt TNG100} and {\tt TNG300} simulations on the stellar-halo mass plane and their best-fit $\log$-$\log$ SHMRs using three representative aperture or outskirt stellar masses discussed in this work.
            From left to right, we show the results for a small aperture (\maper{30}), a large aperture (\maper{100}), and an outskirt (\mout{50}{100}) \mstar{}. In each panel, the orange-red {\tt hexbin} density map represents the distribution of {\tt TNG300} galaxies and the blue scatter points are galaxies from the {\tt TNG100} simulation. The \change{cyan}  vertical lines indicate the peak values of the $M_{\star}$ distributions \changing{and their width reflects the peak difference between {\tt TNG300} and {\tt TNG100}}. We only include galaxies on the right of these lines in the SHMR fitting. We show the best-fit $\log$-$\log$ SHMR of the {\tt TNG300} galaxies using the red dashed line and the {\tt TNG100} relation using the blue dot-dashed line. As the stellar masses on the x-axis span different ranges in the $\log$ scale, it is not easy to quantitatively compare the slopes and scatters of the SHMR. To help the reader visualize the differences among the three panels, we use a solid gray line to highlight an SHMR with slope $a=1.0$. It is easier to see that the SHMR for the outskirt stellar mass has a steeper slope and lower scatter level than the SHMRs of both aperture stellar masses. In the rightmost panel, we compare the halo mass function of {\tt TNG300} simulation (grey line) to the halo mass distributions of the \mstar{}-cut massive galaxy sample ($M_{\star}\geq 10^{11.2} M_{\odot}$; red line) and the sample used to fit the SHMR of the outskirt stellar mass \mout{50}{100} (blue line). We also show a horizontal cyan line to label the \mhalo{} value above which the galaxy samples become $\sim 90$\% complete. 
            The \texttt{Jupyter} notebook for reproducing this figure can be found here: \href{https://github.com/Xuchuyi/HaloMassProxy/blob/main/FigureNotebook/Fig2.ipynb}{\faGithub}.
            }
        \label{fig:fig2}
    \end{figure*}
        
\subsection{Stellar-to-Halo Mass Relation Fitting}
    \label{ssec:shmr}

    The stellar-to-halo mass relation (SHMR) is the most fundamental scaling relation for galaxy formation and galaxy-halo connection (e.g., \citealt{Vale2004MNRAS}; \citealt{Yang2008ApJ}; \citealt{Moster2013MNRAS}). At the high \mstar{} end discussed in this work (\mstar{}$\geq 10^{11.2} M_{\odot}$), \change{the SHMR can be well described by a simple log-log relation (e.g., \citealt{Ziparo2016A&A}; \citealt{Farahi2018MNRAS}; \citealt{Golden-Marx2018ApJ}) using a slope ($a$), an intercept ($b$), and a scatter, which may or may not evolve with mass.}

    When modeling the galaxy-halo connection, \mhalo{} is often seen as the independent variable because it is, in some sense, the more physically fundamental property. In this work, however, we use \mstar{} as the independent variable because we evaluate the performance of different definitions of \mstar{} as proxies for \mhalo{}. First, just as in actual observations, we are dealing with a \mstar{}-complete sample, which makes \mstar{} the more appropriate independent variable from the model-fitting point of view. More importantly, as demonstrated in \citealt{Huang2022}, the scatter of \mhalo{} at fixed \mstar{} is a practical, empirical metric to evaluate an \mhalo{} proxy and can be quantitatively inferred from the data with the help of the galaxy-galaxy lensing method and an N-body simulation. Additionally, a shallower slope implies less evolution in \mhalo{} per unit of \mstar{}, resulting in a smaller additional scatter in \mhalo{} caused by observational uncertainties in \mstar{}. 
    
    Therefore, we consider the \mstar{} whose SHMR has a shallower slope and/or a lower scatter level as a better \mhalo{} proxy.  In this work, we characterize the SHMR scatter using the overall scatter of \logmh{} of all galaxies included during the fitting (\sigmh{}). We ignore the possible variation of the SHMR scatter with the halo mass here. Past work indicated a relatively stable \sigmh{} value at the high-\logmh{} end (e.g., \citealt{Leauthaud2012ApJ}; \citealt{Zu2015MNRAS}; and Fig.8 in \citealt{Wechsler2018ARA&A}). We also do not have a large enough sample of massive haloes to constrain a varying \sigmh{} reliably.

    Although the galaxy sample is \mstar{}-complete above $10^{11.2} M_{\odot}$ \emph{when \mstar{} is defined as the ``total'' stellar mass by the simulation}, this sample is no longer perfectly complete for different aperture or outskirt \mstar{} measurements. For example, a few \mstar{}$< 10^{11.2} M_{\odot}$ galaxies could have extended stellar mass distributions that lead to higher \mout{50}{100} than some galaxies in our sample. We face a similar issue when dealing with observed galaxies, as we often can only afford to estimate the aperture or outskirt \mstar{} on a sample selected based on a default luminosity or stellar mass cut. We decide to take a practical and straightforward approach to deal with this issue. When fitting the SHMR for a given \mstar{} measurement, we only include galaxies above the \emph{peak} of the \mstar{} distribution\footnote{We use the histogram of the stellar mass in 30(20) bins to describe the stellar mass distribution for {\tt TNG300}({\tt TNG100},{\tt Illustris}). The choice of bin numbers does not affect the key results.} (labeled as the vertical lines in Fig.~\ref{fig:fig2}). Although this simple approach does not guarantee the completeness of the sample, it should greatly alleviate its impact on our main conclusions.
    
    In addition, we notice that a small fraction of simulated massive galaxies show unrealistically low \ins{} or \exs{} stellar masses. For example, we find the following counts of galaxies with $M_{\rm ins} < 10^{10} M_{\odot}$ and $M_{\rm exs} < 10^{10.5} M_{\odot}$ in each simulation: in {\tt TNG100}, 0/235 and 1/235; in {\tt TNG300}, 7/2713 and 12/2713; and in {\tt Illustris}, 3/339 and 24/339, respectively. After checking the 2-D stellar mass maps of these outliers, we conclude that although they could represent some sporadic cases, it is more likely that such low \ins{} or \exs{} values are due to problems when assigning stellar particles to the central galaxies. \change{Thus, we excluded these outliers from the SHMR fitting, even though they would not impact the key results.}

    For each of the stellar mass measurements, we use the linear regression algorithm \change{in {\tt scipy} to fit a log-log SHMR above the peak stellar mass value:}
    \[\log_{10}[M_{\rm halo}/M_\odot]=a \times \log_{10}[M_{\star}/M_\odot]+b\]
    In Fig.~\ref{fig:fig2}\footnote{The difference between the {\tt TNG300} and {\tt TNG100} lines indicates that, at a fixed halo mass, massive galaxies in {\tt TNG100} exhibit a higher stellar mass. This discrepancy arises from the resolution differences between {\tt TNG300} and {\tt TNG100} (see Fig. A1 in \citealt{Pillepich2018MNRAS}).}, we show the best-fitting mean SHMRs for a small aperture (\maper{30}), a large aperture (\maper{100}), and an outskirt (\mout{50}{100}) stellar mass for the {\tt TNG100} and {\tt TNG300} samples. Using these log-log SHMRs that fit the best, we then estimate the \sigmh{} levels by directly calculating the standard deviations of the distributions for $\{\log[M_{\rm halo}/M_\odot]-a^\star\log[M_\star/M_\odot]\}$. We also bootstrapped the sample 8000 times to estimate the uncertainties of slope, intercept, and \sigmh{}. 

    In observations, both the uncertainties of photometry and the mass-to-light ratio ($M/L$) affect the estimation of \mstar{}; therefore, they also impact the best-fit SHMR. To study this effect, we assume that these uncertainties can be statistically described by a Gaussian distribution in the \logms{} space whose variance is $\epsilon^2_{\log{M_{\star}}}$. For all aperture stellar masses, we assume $\epsilon_{\log{M_{\star}}} = 0.1$ dex. For the outskirt stellar masses, since the photometric uncertainties in the low surface brightness outskirts are typically higher (\citealt{Huang2018}), we assume $\epsilon_{\log{M_{\star}}} = 0.15$ dex. Based on these assumptions, we resample each stellar mass value 1500 times following a $\mathcal{N}(\log_{10} M_{\star}, \epsilon^2_{\log{M_{\star}}})$ to generate a series of ``mock observations'' of our samples. In these samples, we repeat the SHMR fitting procedures and estimate the slopes and scatters of the SHMRs along with their statistical uncertainties. These results will help us more realistically evaluate the different \mhalo{} proxies.

    In addition to all the aperture and outskirt \mstar{} measurements mentioned above, we also perform the same $\log$-$\log$ SHMR fitting for the \mstar{} of the \ins{} and \exs{} components. Usually, in both hydro-simulations (e.g., \citealt{Pillepich2018MNRAS}) and \change{semi-empirical models (e.g., \citealt{Behroozi2019MNRAS}; \citealt{Bradshaw2020MNRAS, Huang2020MNRAS})}, the \exs{} stars often show a tighter correlation with \mhalo{} than with \ins{} and the total \mstar{} due to their close physical connection to the halo assembly history. Although it is still challenging to decompose \exs{} stars reliably in observations (but see \citealt{Zhu2022}), the SHMR of \exs{} \mstar{} can serve as a baseline to compare with other more practical \mhalo{} proxies.


\section{Results} 
    \label{sec:results}

    In this section, we present our main results. We start in \S\ref{ssec:res_empirical} with the evaluation of different \mhalo{} proxies based on their best-fit $\log$-$\log$ SHMR slope and scatter values under different projections and in other simulations. In \S\ref{ssec:topN}, we present the results of the Top-$N$ tests and compare the \sigmh{} values of different halo mass proxies using different binning strategies.

    \begin{figure*}[htb]
        \centering
        \includegraphics[width=1.1\linewidth,trim=3cm 0 0 0]{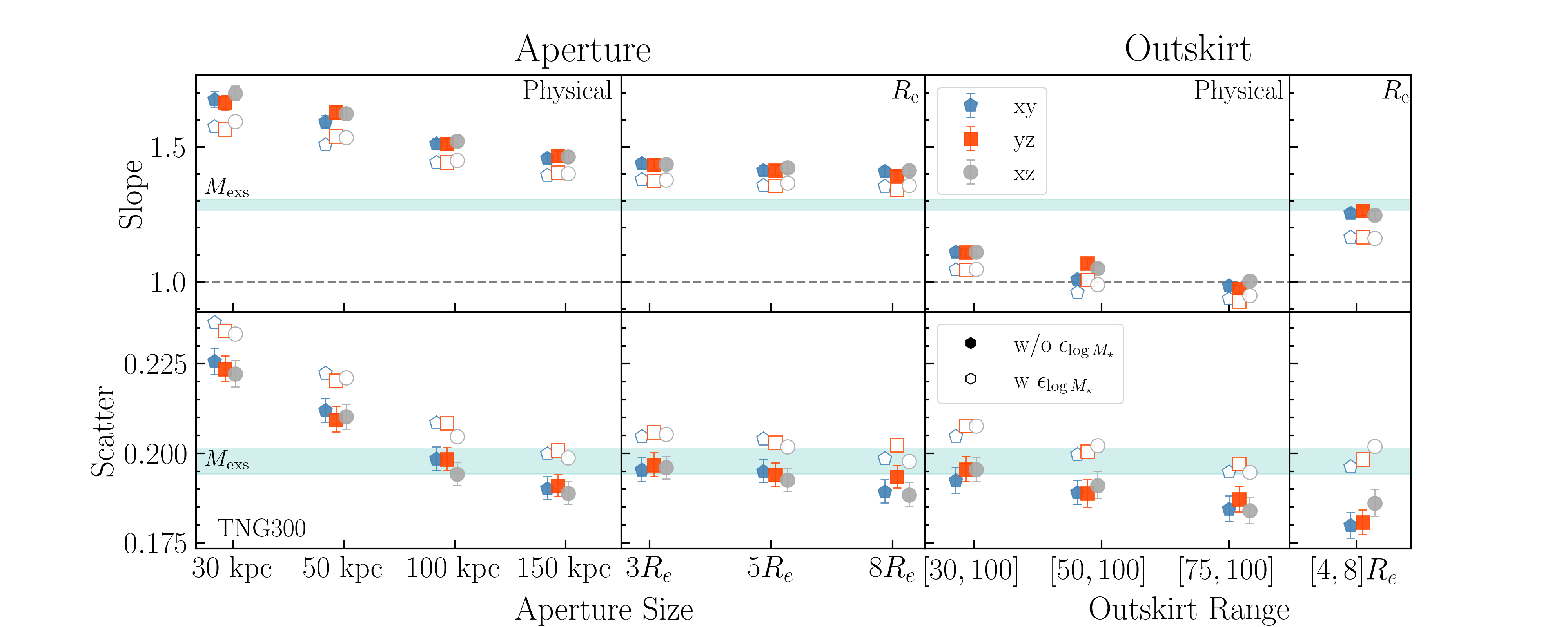}
        \caption{
            \change{Slopes} (top panels) and \changing{scatters (\sigmh{}, bottom panels)} values of the $\log$-$\log$ SHMRs for different \mhalo{} proxies using the 2-D stellar mass distributions of massive galaxies in {\tt TNG300} using three orthogonal projections ({\tt XY} in blue; {\tt YZ} in orange; and {\tt XZ} in gray; we shift the symbols horizontally to improve visibility). We divide the figure into four sections according to the types of stellar mass measurements. From left to right, different apertures and outskirts are shown.  We consider the apertures within 30, 50, 100, and 150 kpc and within 3, 5, and 8 $R_{\rm e}$. The outskirts are evaluated at the [30,100], [50, 100], and [75, 100] kpc bins and at the [4,8]$R_{\rm e}$ bin. For aperture \mstar{}, we label the aperture size on the X-axis. For the outskirt stellar mass, we show their inner and outer boundary values in square brackets. In all panels, solid symbols represent direct stellar mass measurements using the 2-D stellar mass maps. Their error bars show the $1\sigma$ confidence interval from the bootstrap resampling. The open symbols represent SHMRs with realistic statistical uncertainties in the stellar mass measurements. We assume that the stellar mass measurements follow a Gaussian distribution and take multiple random draws to estimate the mean values of the slope and scatter along with their uncertainties. We also show the slope and scatter values for the SHMR of \exs{} \mstar{} without additional uncertainties using the horizontal cyan bar, whose width represents the 1-$\sigma$ ranges from the bootstrap resampling same as the error bar of points. In the panels for the SHMR slopes, we also use a black dashed line to highlight the slope $=1$ value. Since we chose \mstar{} as the independent variable when fitting the SHMR, a better \mhalo{} proxy would have a lower slope value and/or a lower scatter value in this figure. By this standard, the outskirt \mstar{} with a $> 50$ kpc inner boundary is the best \mhalo{} proxy in our tests. Furthermore, different projections leave little mark on the slope and scatter values of the {\tt TNG300} sample.
            The \texttt{Jupyter} notebook for reproducing this figure can be found here: \href{https://github.com/Xuchuyi/HaloMassProxy/blob/main/FigureNotebook/Fig3.ipynb}{\faGithub}.
            } 
        \label{fig:fig3}
    \end{figure*}

    \begin{figure*}[htb]
        \centering
        \includegraphics[width=1.1\linewidth,trim=3cm 0 0 0]{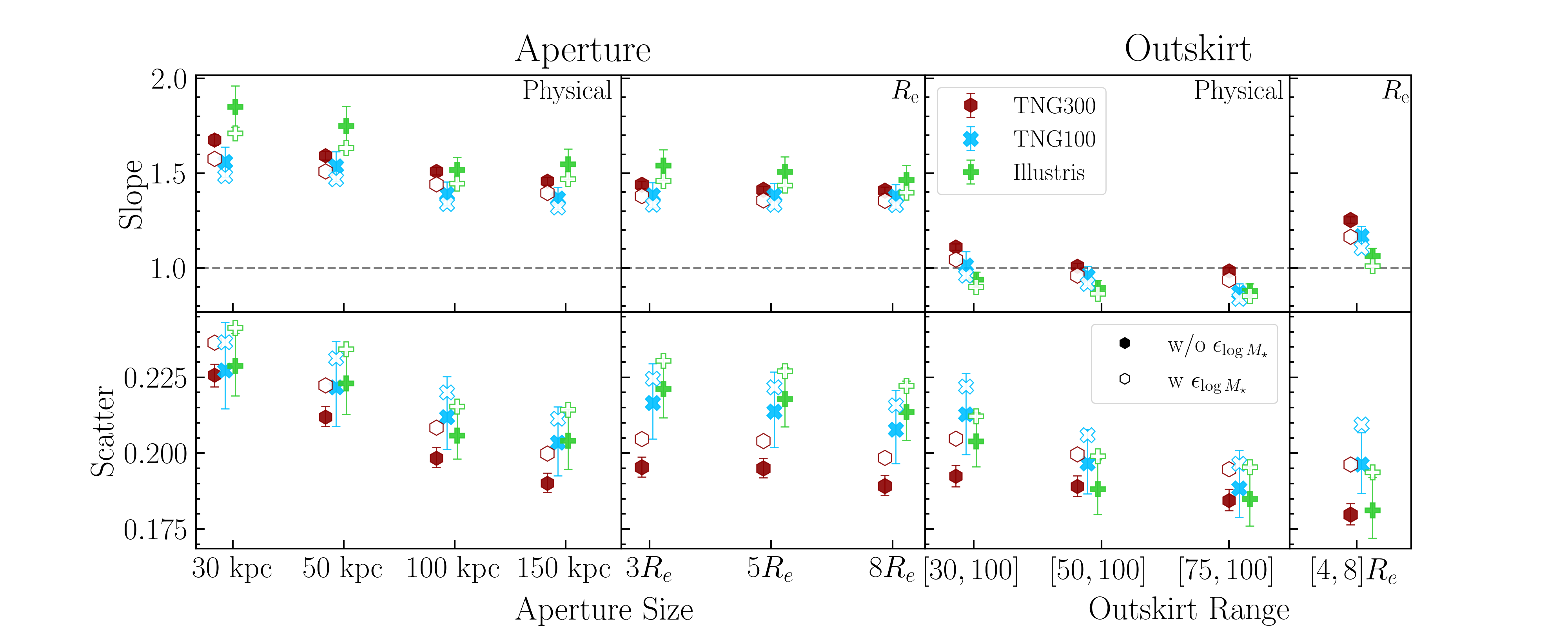}
        \caption{
            Slopes (top panels) and scatters (\sigmh{}, bottom panels) of the SHMRs for massive galaxies from different hydro-simulations ({\tt TNG300}, {\tt TNG100}, and {\tt Illustris}). The format of this figure is identical to that in Figure \ref{fig:fig3}. This figure highlights the differences in the SHMRs of massive galaxies caused by the mass resolution (e.g., {\tt TNG300} vs. {\tt TNG100}) or the physical recipes adopted in the simulation (e.g., {\tt TNG100} vs. {\tt Illustris}). 
            The \texttt{Jupyter} notebook for reproducing this figure can be found here: \href{https://github.com/Xuchuyi/HaloMassProxy/blob/main/FigureNotebook/Fig4.ipynb}{\faGithub}.
            }
    \label{fig:fig4}
    \end{figure*}
    
\subsection{Outskirt $M_\star$ as Halo Mass Proxy}
    \label{ssec:res_empirical}

    Following the methods described in \S\ref{ssec:shmr}, we now evaluate the performance of different stellar masses as halo mass proxies based on the slopes and scatters of their SHMRs.
    In Fig.~\ref{fig:fig3}, we first summarize the key results for massive galaxies in the {\tt TNG300} simulations. The large volume of {\tt TNG300} translates into a larger sample size of massive galaxies, making it more appropriate for statistical evaluations of \mhalo{} proxies. As a reminder from \S\ref{ssec:shmr}, based on observational perspectives, we prefer a \mhalo{} proxy with \emph{lower} slope (since we choose to use \mstar{} as the independent variable) and \emph{lower} \sigmh{} values. In Fig.~\ref{fig:fig3}, we use the results of the \exs{} \mstar{} (cyan-shaded regions) and the slope $a=1$ line (black, dashed) as references. 

    First, when using the physical unit (kpc), we find that the performance of an aperture \mstar{} as a \mhalo{} proxy gradually improves with its aperture size, from 30 to 150 kpc, based on its slope and \sigmh{} values. This trend is qualitatively consistent with the findings in \citet{Huang2022} using $z<0.5$ massive galaxies from the HSC survey. \changing{And a similar trend in the slope of SHMR has also been observed in other studies (e.g. \citealt{Golden-Marx2019ApJ}; \citealt{Golden-Marx2023MNRAS}; \citealt{Golden-Marx2025MNRAS})}. Our result confirms that \change{\emph{the stellar mass measurements based on a small aperture or a shallow image tend to miss the extended outskirts of massive galaxies and cannot be a good \mhalo{} proxy or indicator}}. When the aperture size is larger than 100 kpc, the \sigmh{} value of the aperture \mstar{} is comparable to the scatter for \exs{} \mstar{} at $\sim 0.20$ dex. However, the best slope value of an aperture \mstar{} is around 1.5, higher than the $\sim 1.3$ value for \exs{} stars, suggesting that the total \exs{} \mstar{} still outperforms any large aperture \mstar{} for available imaging data. Using $R_{\rm e}$ as a reference, we find that the performance of all the large apertures \mstar{} we evaluated, from $3 \times$ to $8 \times R_{\rm e}$, is very similar to that of \maper{150} and does not vary with aperture size. The lack of dependence on aperture size suggests that $R_{\rm e}$ already carries useful information on halo mass. \change{In observations}, it is up for debate whether the ``size'' of an early-type galaxy depends on the halo mass \emph{at fixed stellar mass} (e.g., \citealt{Charlton2017MNRAS}; \citealt{Sonnenfeld2022A&A}) since the definition of the ``size'' and the method to estimate it from the data could affect the conclusion. Here, in {\tt TNG300}, we demonstrate that \emph{if we can accurately measure the effective radius}, \maperre{3} can match the performance of a large aperture \mstar{} such as \maper{150}.

    Moving to the outskirt \mstar{}, in Fig.~\ref{fig:fig3}, we find that the three measurements defined within physical radii $>30$ kpc outperform the best aperture \mstar{} we evaluated. The \sigmh{} values of the outskirt \mstar{} gradually decrease as the inner boundary increases from 30 ($\sim 0.195$ dex) to 75 kpc ($\sim 0.185$ dex), which is better than \exs{} \mstar{} ($\sim 0.20$ dex). More importantly, the most significant improvement in the outskirt \mstar{} as a \mhalo{} proxy comes from \emph{the slope of its SHMR}, which is not only significantly lower than the best aperture \mstar{} ($\sim 1.5$) and \exs{} \mstar{} ($\sim 1.3$), but also approaches 1.0. The slope values of the outskirt \mstar{} also decreases slightly from $\sim 1.1$ of \mout{30}{100} to $\sim 1.0$ of \mout{75}{100}. When switching to the outskirt \mstar{} defined by $R_{\rm e}$ (\moutre{4}{8}), we notice that, while the \sigmh{} value is very low ($\sim 0.18$ dex), the slope of its SHMR ($\sim 1.25$) is on par with the \exs{} \mstar{} but higher than that of the outskirt peers with the physical unit. Although we only highlight the results for \moutre{4}{8}, we tested further binnings, obtaining consistent results.

    As we explore different \mstar{}-based \mhalo{} proxies, we know that the simulation's resolution and the adopted physical recipes could affect the galaxy-halo connection and structures of massive galaxies (\citealt{Pillepich2018MNRAS, Ardila2021MNRAS}). To examine the robustness of the results mentioned above, we further compare the slope and \sigmh{} values of the SHMRs for different aperture and outskirt \mstar{} in three hydro-simulations: \illustris{}, {\tt TNG100} and {\tt TNG300}. Among these three, {\tt TNG300} has a lower mass resolution than \illustris{} and {\tt TNG100}. The original \illustris{} simulation is different from the {\tt TNG} series in several critical aspects of galaxy formation physics. We present the results of this comparison in Fig.~\ref{fig:fig4}. We find that qualitative evaluations of different \mhalo{} proxies based on their SHMRs are consistent among the three simulations. Compared to {\tt TNG300}, simulations with higher mass resolution show slightly larger \sigmh{} values. The much smaller volumes of the \illustris{} and {\tt TNG100} simulations could play a role in this, but it could also be the genuine effect of the lower resolution (see Leidig et al. in prep. for details). Meanwhile, we find that massive galaxies in \illustris{} tend to have higher slope values for aperture \mstar{} (especially for smaller apertures such as \maper{30} \& \maper{50}) and lower slope values for the outskirt \mstar{} compared to galaxies in {\tt TNG} simulations. Regardless of these subtle differences, Fig.~\ref{fig:fig4} confirms that the outskirt stellar mass (e.g., \mout{50}{100}) is an excellent \mstar{}-based \mhalo{} proxy in simulations with different resolution and galaxy formation physics. Although more simulations are worth exploring in the future, this conclusion is encouraging. 

    Based on the results shown in Fig.~\ref{fig:fig3} and Fig.~\ref{fig:fig4}, we conclude that, for low-redshift massive galaxies in {\tt TNG300}, the outskirt stellar mass defined using fixed physical aperture sizes (e.g., between 50 and 100 kpc) is the most promising \mhalo{} proxy. In Fig.~\ref{fig:fig3}, we also examine the influence of three different projections and the impact of statistical uncertainties on \mstar{}. All conclusions remain the same. Using the stellar mass maps from the three orthogonal projections of the simulation, we find a little difference in the slope and \sigmh{} values of different SHMRs explored here. But this is consistent with each other within the estimated $1\sigma$ uncertainties, demonstrating the robustness of the conclusion. However, it does not suggest that the orientation of a massive galaxy's stellar halo relative to the observer might not affect the stellar mass measurements in 2-D. We will discuss this in \S\ref{ssec:proj}. 

    As indicated by the open symbols in Fig.~\ref{fig:fig3} and Fig.~\ref{fig:fig4}, the statistical uncertainty of \mstar{} makes the slope of the SHMR moderately shallower and the value of \sigmh{} higher for all the \mhalo{} proxies evaluated while leaving all main conclusions qualitatively unchanged and consistent with the results of \citealt{Huang2022}.


    \begin{figure*}
        \centering
        \includegraphics[width=1.1\linewidth,trim=2cm 0 0 0]{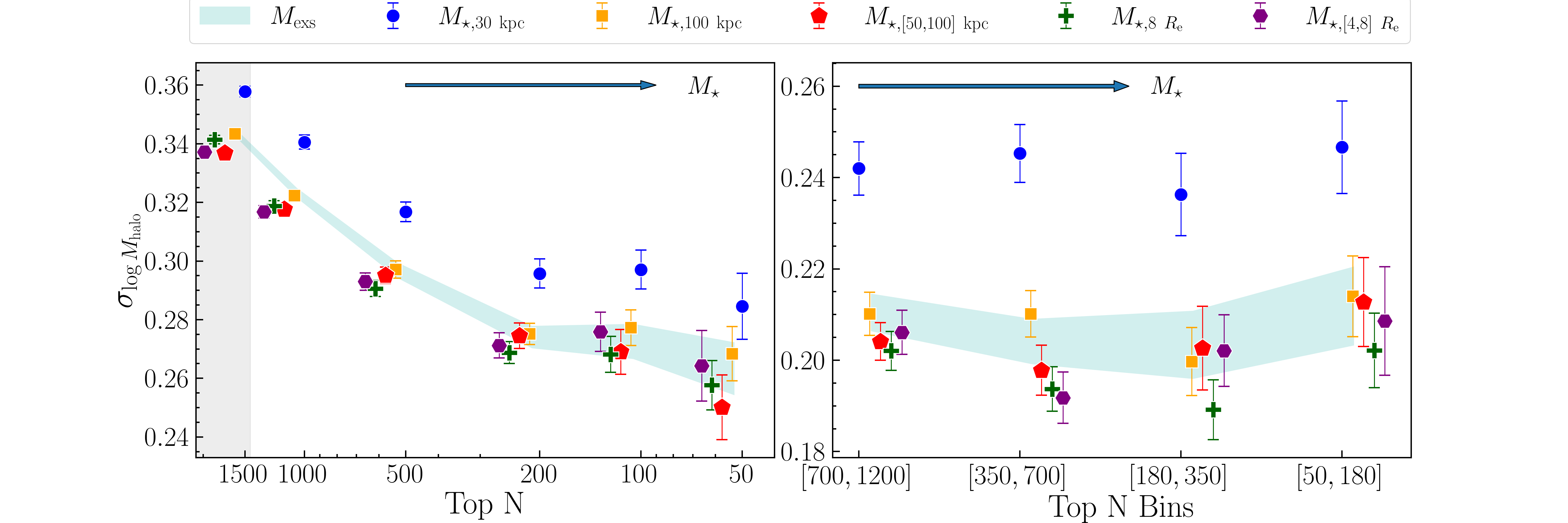}
        \caption{
            {\bf Left panel:} The scatters of halo mass ($\sigma_{\log M_{\rm halo}}$) in different ``Top--$N$'' bins of {\tt TNG300} galaxies based on a series of halo mass proxies. 
            The X-axis represents the ``$N$'' values of the ``Top--$N$'' bins -- the number of galaxies included in each bin. 
            \change{The mean stellar mass} of the ``Top--$N$'' bin monotonically increases from the left to the right.
            For the $\sigma_{\log M_{\rm halo}}$ values shown on the Y-axis, a lower value indicates that the property examined is a better halo mass proxy than the one with a higher scatter value.
            We include both the aperture and the outskirt stellar masses defined in units of kpc or $R_{\rm e}$ as halo mass proxies here.
            To improve visibility, we shift the symbols horizontally for the same ``Top--$N$'' bins. 
            The error bars show the $1\sigma$ confidence interval of the uncertainties in stellar mass measurements.
            We also use the cyan-shaded region to represent the $\sigma_{\log M_{\rm halo}}$ error bars of the Top-$N$ samples selected by their true \exs{} stellar masses.
            The outskirt $M_{\star}$ defined in either kpc or $R_{\rm e}$ significantly outperforms the aperture ones and is on par with the \exs{} stellar mass.
            {\bf Right panel:} Similar to the left panel, but using the samples based on consecutive bins of $N$. 
            The X-axis shows each bin's lower and upper indices, where $[50, 180]$ represents the sample from the top 50 to the top 180 based on the values of the evaluated halo mass proxies.
            To highlight the relative differences among the different proxies, we compress the $\sigma_{\log M_{\rm halo}}$ range represented by the Y-axis in the right panel.
            The right panel also lends support to the fact that the outskirt $M_{\star}$ is a good $M_{\rm halo}$ proxy, consistent with \citet{Huang2022}.
            The \texttt{Jupyter} notebook for reproducing this figure can be found here: \href{https://github.com/Xuchuyi/HaloMassProxy/blob/main/FigureNotebook/Fig5.ipynb}{\faGithub}.
            }
        \label{fig:fig5}
    \end{figure*}

\subsection{Top-$N$ test}
    \label{ssec:topN}

    Although we have established the conclusion that the outskirt stellar masses such as \mout{50}{100} are excellent \mhalo{} proxies based on the slope and scatter of $\log$-$\log$ SHMR, it is challenging to derive them in observation in a direct manner, though indirect inference is possible\footnote{\change{Note that the SHMR scatter value could be inferred from observation when relying on a secondary halo mass proxy calibrated by weak lensing data, e.g., SZ effect in \citet{Golden-Marx2023MNRAS}.}}. In \citet{Huang2022}, the authors proposed the \topn{} test, an empirical method to evaluate different \mhalo{}-proxies with the help of an N-body simulation and galaxy-galaxy lensing measurements. We refer the reader to \citet{Huang2022} for a detailed description of the \topn{} test, but its logic is straightforward. For two different observables as candidates for \mhalo{} proxies, one can rank their values in descending order\footnote{This assumes the value of the observable \emph{positively} correlates with the halo mass. In the case of a \emph{negatively} correlated property, one could rank the property values in ascending order} and take the first $N$ objects to form two samples (``Top $N$''). The average value and scatter of the halo masses of these two samples provide an objective metric to evaluate them as \mhalo{} proxies. Given the form of the halo mass function at high-\mhalo{} end (halo number density steeply decreases toward higher \mhalo{} value) and assuming a $\log$-$\log$ scaling relation, the better \mhalo{} proxy should have \emph{higher} average \mhalo{} and \emph{lower} scatter of \mhalo{} in the same Top-$N$ sample. \change{In observations, the stacked galaxy-galaxy lensing profile (e.g., the excess surface density, or \dsigma{} profile) of the Top-$N$ sample qualitatively reflects the average \mhalo{} and the scatter (\citealt{Huang2022})}, \change{using the massive halos in an $N$-body simulation} with the appropriate volume, one can further estimate the \mhalo{} scatter within the sample. In hydro-simulations like \tng{}, we can directly calculate the \mhalo{} scatter values to compare different stellar mass measurements.

    In this work, we examine the results of the empirical tests \topn{} as a qualitative approach to compare with observations in Fig.~\ref{fig:fig5}. We only use {\tt TNG300} galaxies for the Top-$N$ test, as this statistical test requires a sufficiently large sample of massive halos. First, for each rank-ordered aperture or outskirt stellar mass measurement, we select the top $N={50, 100, 200, 500, 1000, 1500}$ galaxies to create a series of Top-$N$ samples. We then estimate the scatter of \mhalo{} in the form of \sigmh{}. These scatter values can help us assess the performance of different \mhalo{} proxies above different \mhalo{} thresholds. \change{For {\tt TNG300}, these number density thresholds translate into \logmh{}$=$[14.3,14.1,14.0,13.7,13.5,13.3]}

    In addition to the above Top-$N$ samples, we also follow the strategy in \citet{Huang2022} to define Top-$N$ samples based on consecutive Top-$N$ bins between two different $N$ to evaluate a \mhalo{} proxy within different \mhalo{} ranges. As demonstrated in \citet{Huang2022}, while the scatter of \mhalo{} in each of these consecutive Top-$N$ bins is a composite view of the slope of SHMR and the scatter of \mstar{} at fixed \mhalo{} (\sigms{}), it provides a simple approach to describe the SHMR empirically without invoking any strong assumptions. Also, from a practical point of view, the \sigmh{} value of a sample defined by an observable is what we care the most about for applications in cosmology or galaxy-halo connection modeling. This work defines four Top-$N$ bins for each \mstar{} measurement: $[50,180]$, $[180,350]$, $[350,700]$, $[700,1200]$ as shown in the right panel of Fig.~\ref{fig:fig5}. We choose these bins based on the total \mstar{} distribution of the {\tt TNG300} sample so that each bin approximately corresponds to $\sim0.15$ dex in \logms{}.

    As described above in \S\ref{ssec:shmr}, we also consider the statistical uncertainties of different aperture and outskirt stellar mass measurements. Following the same strategy, we randomly draw from these distributions 1000 times to calculate the mean Top-$N$ scatter values and the associated statistical uncertainties.

    Therefore, we design two sets of samples following the strategy of \citet{Huang2022}: the simple ``Top-$N$'' samples (left panel in Fig.~\ref{fig:fig5}) and consecutive ``Top-$N$'' bins (right panel). For each halo mass proxy, we rank-order the sample based on their values, select them into these two types of bins, and calculate the scatter of halo mass $\sigma_{\log{M_{\rm halo}}}$ within each bin. 

    \changing{Putting} the results for both selections together, we find that:
    \begin{itemize}
    
        \item Stellar mass within a 30 kpc aperture (about 1.5-3.0 $\times R_{\rm e}$ for galaxies in our sample) consistently underperforms as a halo mass proxy compared to large-aperture or outskirt stellar masses in \emph{all} bins, consistent with \citet{Huang2022} (see their Figure 5) . This emphasizes the importance of carefully selecting galaxy stellar mass definitions when studying galaxy-halo connections of massive galaxies, especially when dealing with the stellar mass based on the conventional photometric pipeline used in modern imaging surveys (such as HSC-SSP program, \citealt{Bosch2018PASJ}) or hydro-simulation outputs. 

        \item When using fixed physical sizes to define the aperture or outskirt, we see that the outskirt stellar mass between 50 to 100 kpc outperforms the stellar mass within a 100-kpc aperture in the \topn{} tests, especially at the higher cumulative number density (or lower halo \& stellar mass) regime, which is, again, consistent with \citet{Huang2022} (see their Figure 5 \& 6). Although the smaller \tng{} sample size limits statistical significance, we still find evidence that \mout{50}{100} outperforms total \exs{} stellar mass when $N > 350$ in \topn{} tests.

        \item When using $R_{\rm e}$ to define the aperture or outskirt region, we notice that the large-aperture stellar mass within $8 \times R_{\rm e}$ and the outskirt stellar mass between 4 and $8 \times R_{\rm e}$ slightly but consistently outperform the \mout{50}{100}. The better performance of $R_{\rm e}$-based \mstar{} is, in principle, attractive, as it could bypass the semi-arbitrary choices of physical sizes (e.g., 30 kpc, between 50 to 100 kpc). More importantly, $R_{\rm e}$-based \mstar{} can adapt to the mass-size relations of galaxies; hence, we could apply them in less massive galaxies or even beyond the early-type populations. In Appendix E of \citet{Huang2022}, the authors also explored several $R_{\rm e}$-based stellar masses out to $6 \times R_{\rm e}$, but found none with better performance than \mout{50}{100}. \change{This difference may reflect the intrinsic differences between the mass-size relations of massive galaxies from HSC and in \tng{} simulation and the inherent difficulty in measuring the $R_{\rm e}$ observationally.} Yet, with the upcoming deep imaging surveys, it is worth exploring the potential of $R_{\rm e}$-based \mstar{} as a halo mass proxy soon. 

        \item When using the Top-$N$ bins (right panel of Fig.~\ref{fig:fig5}), we notice that the $\sigma_{\log{M_{\rm halo}}}$ values roughly stay the same and show no clear trend with the number density ranges. Given that these bins correspond to a relatively narrow stellar mass range, the lack of trend implies that the SHMR at $\log[M_\star/M_\odot]\gtrsim11.5$ shows little variation in their slopes and intrinsic scatters. However, this behavior is different from the results in \citet{Huang2022} (see Figure 5 \& 6), where the scatter values steadily increase toward the higher number density (lower halo mass) end. Also, compared to the $\sigma_{\log{M_{\rm halo}}}$ values reported here ($<0.26$ dex for the consecutive bins), the HSC results show significantly higher scatter values ($> 0.3$ dex). Such qualitative and quantitative differences could relate to the different approaches to estimate $\sigma_{\log{M_{\rm halo}}}$ values. In \citet{Huang2022}, the authors derive these scatter values for the HSC massive galaxies through the stacked galaxy-galaxy lensing profiles and with the help of N-body simulation \& semi-empirical models. Compared to the ``true'' halo mass scatter calculated here, the scatters from observation may include systematic issues and modeling uncertainties. With the availability of more realistic mock images (e.g., \citealt{Bottrell2024MNRAS}) and the arrival of even larger hydro-simulations, we will seek more realistic evaluations of \mhalo{} proxies that include systematic issues in observation. 
        
    \end{itemize}


\section{Discussion} 
    \label{sec:discussions}
    

\subsection{The connection between \exsitu{} stellar mass and halo mass}
\label{ssec:exsscat}

    In the two-phase formation scenario, the traditional definition of the \exs{} component includes the stars in the satellite galaxies accreted into the more massive central galaxy throughout its \emph{entire} life. By definition, these \exs{} stars hold a physical connection to \emph{all} the subhalos in the parent halo's assembly history. As these subhalos can contribute significantly to the halo mass built up of today's massive dark matter halos (e.g., \citealt{Zhao2003MNRAS}; \citealt{Genel2010ApJ}), it is natural to expect that the total amount of \exs{} stars, as a summary of this powerful physical connection, should be an excellent halo mass proxy. In both hydro-simulations (e.g., \citealt{Pillepich2018MNRAS}) and semi-empirical models (e.g., \citealt{Bradshaw2020MNRAS}), \exs{} stellar mass indeed often correlates better with halo mass than the \ins{} component or the total stellar mass (within a spherical region centered on the galaxy). In \citet{Huang2022}, the authors speculated that the outskirt stellar masses' good performance as a halo mass proxy might be because it measures the ``historic richness'' and represents a fraction of the satellites accreted in the past. Under this logic, we would naively expect the total \exs{} stellar mass to outperform outskirt stellar mass such as \mout{50}{100}, as the latter only includes a fraction of the \exs{} stars (plus the ``contamination'' of a small fraction of \ins{} stars). 

\begin{figure}
        \centering
        \includegraphics[width=1.1\linewidth,trim=0.5cm 0 0 0]{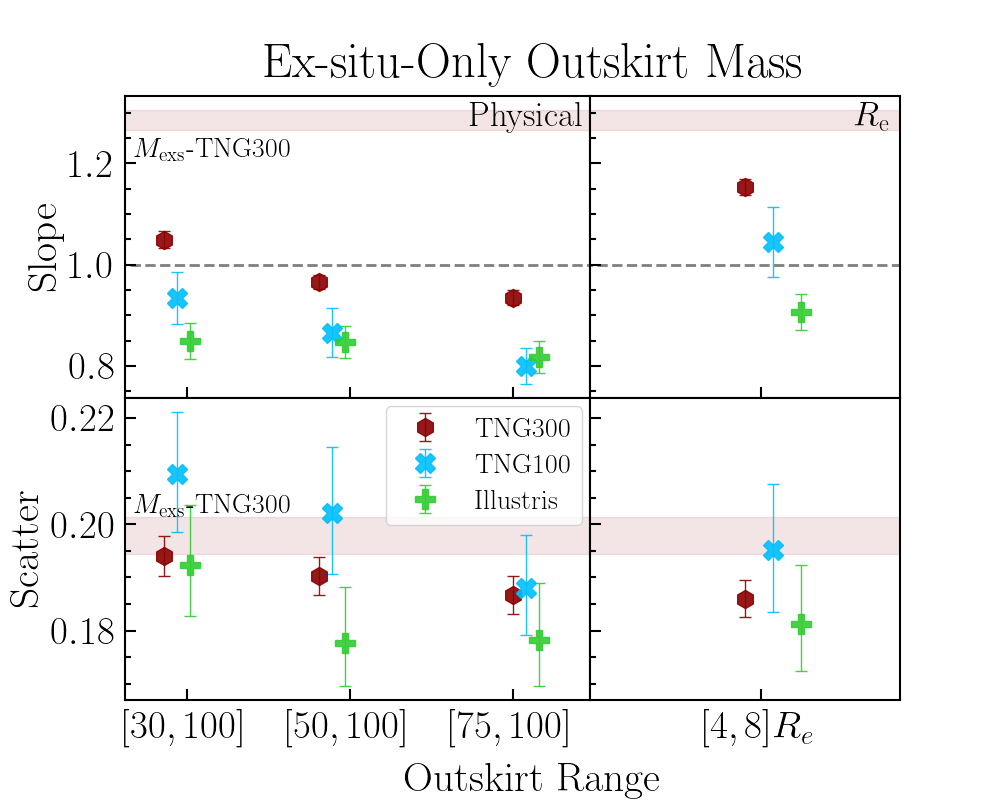}
        \caption{
            A summary of the slopes and scatters of the SHMRs for the outskirt \exs{} stellar masses using all three simulations explored in this work.
            The format is the same with the right two sections of Fig.~\ref{fig:fig4}.
            The red-shaded regions represent the total \exs{} stellar mass values in {\tt TNG300}.
            Note that we still use the $R_{\rm e}$ of the total stellar mass distribution in this figure, \emph{not} the effective radius of the \exs{} component. 
            While the \exs{}-only outskirt masses are excellent halo mass proxies, they show no significant improvement when compared to the outskirt masses based on the total stellar mass distributions (see Fig.~\ref{fig:fig4}).
            This result is consistent with the \exs{} component dominating the outskirts of massive galaxies.
            Meanwhile, the fact that these outskirt \exs{} masses have better performance than the total \exs{} mass suggests that the inner and outer \exs{} components may have different relations with the halo assembly history.
            The \texttt{Jupyter} notebook for reproducing this figure can be found here: \href{https://github.com/Xuchuyi/HaloMassProxy/blob/main/FigureNotebook/Fig6B.ipynb}{\faGithub}.
            }
        \label{fig:fig6}
    \end{figure}

    However, as shown in Fig.~\ref{fig:fig3} and Fig.~\ref{fig:fig5}, we find that this is not the case for \changing{the} \illustris{} and \tng{} simulations: \change{the outskirt stellar masses explored in this work consistently show comparable or better performance as halo mass proxies than the \exs{} stellar mass.} Interestingly, while Fig.~\ref{fig:fig3} and Fig.~\ref{fig:fig5} already show that the outskirt stellar masses show better performance as halo mass \change{proxies} than the total \exs{} stellar mass, we want to emphasize that even the \exs{} version of the outskirt stellar masses outperform the total \exs{} stellar mass in all three simulations using Fig.~\ref{fig:fig6}. This finding not only confirms that outskirt stellar mass is a good halo mass proxy, but it also reveals a significantly deeper connection of the outer part of the \exs{} stars to halo growth compared to both the entire \exs{} component and the inner \exs{} component, at least in hydro-simulations. 

    While this seemingly surprising result still begs for more confirmation and investigation, we suggest that it might be rooted in 1. the connection between the halo assembly and the stellar halo build-up; 2. the unsatisfying definition of \exs{} component. First, in N-body simulations, massive dark matter halos also grow in two phases: the early, fast-accretion phase defines the halo's potential well, and a second slow-accretion phase that primarily piles mass in the halo outskirt without altering the potential well but gradually increases the halo's concentration (e.g., \citealt{Zhao2003MNRAS}). Although it is still too early to say that the two-phase assembly of massive galaxies mirrors this two-phase halo growth, today's mass of a massive dark matter halo might have a stronger connection to the slow-accretion phase, where minor mergers of subhalos (and satellite galaxies) dominate the merger history and help build up the outer stellar halos around massive galaxies. As \citealt{RodriguezGomez2016MNRAS} and \citealt{Montenegro-Taborda2023} showed, \exs{} stars from minor mergers prefer the outer regions in hydro-simulations while major mergers are the main ones responsible for the ex-situ stellar populations residing in the innermost regions of massive galaxies. Secondly, the current definition of \exs{} component is a purely theoretical one based on simulations (e.g., \citealt{RodriguezGomez2016MNRAS}) and traces the entire galaxy evolution up to the extreme high-redshift Universe (e.g., $z>4$). This definition sometimes creates situations where the \exs{} component completely dominates a massive galaxy's stellar mass distribution, even in the central region, making observational decomposition virtually impossible. \change{And, the \exs{} component resulting from a gas-rich, high-redshift major merger might be indistinguishable from the corresponding \ins{} component in terms of stellar density profiles, making it more challenging to realize this definition of \exs{} component in observations.} More importantly, such a definition also delivers an oversimplified, if not biased, characterization of a galaxy's mass accretion history. For instance, under the current definition, when a major merger with a 1:0.99 mass ratio happens, \emph{all} the stars from the slightly less massive galaxy would join the \exs{} component of the descendent, which is not perfectly physical: the ``satellite'' galaxy here might have a similar halo mass (assembly history) and a similar amount of its own \exs{} component with the ``central''. Assuming they have the same halo mass, such a merger event would double the halo mass but increase the \exs{} fraction more significantly (depending on the original \exs{} fraction of the ``more massive central''). While this is hypothetical, it demonstrates how the \exs{} definition could lead to the ``degraded'' performance of the total \exs{} stellar mass as a halo mass proxy.
    
    Therefore, the reasons to \emph{not} recommend \exs{} stellar mass as a halo mass proxy include: 1. practically speaking, it is unclear that we can ``decompose'' the \exs{} stars in a massive galaxy using images or other observations (also see \citealt{Zhu2022}; \citealt{Angeloudi2024NatAs}); 2. even if we can measure the \exs{} stellar mass accurately, the current definition of the \exs{} component could make it a less appealing proxy of today's halo mass. That said, we acknowledge that the definition of our outskirt stellar mass is still purely empirical and too arbitrary. We should further pursue the detailed connection between the halo assembly history and the spatially-resolved growth of massive galaxies' stellar halo after considering specific secondary halo properties, such as the halo concentration. 
    

    \begin{figure*}
        \centering
        \includegraphics[width=1.05\linewidth,trim=2cm 0 0 0]{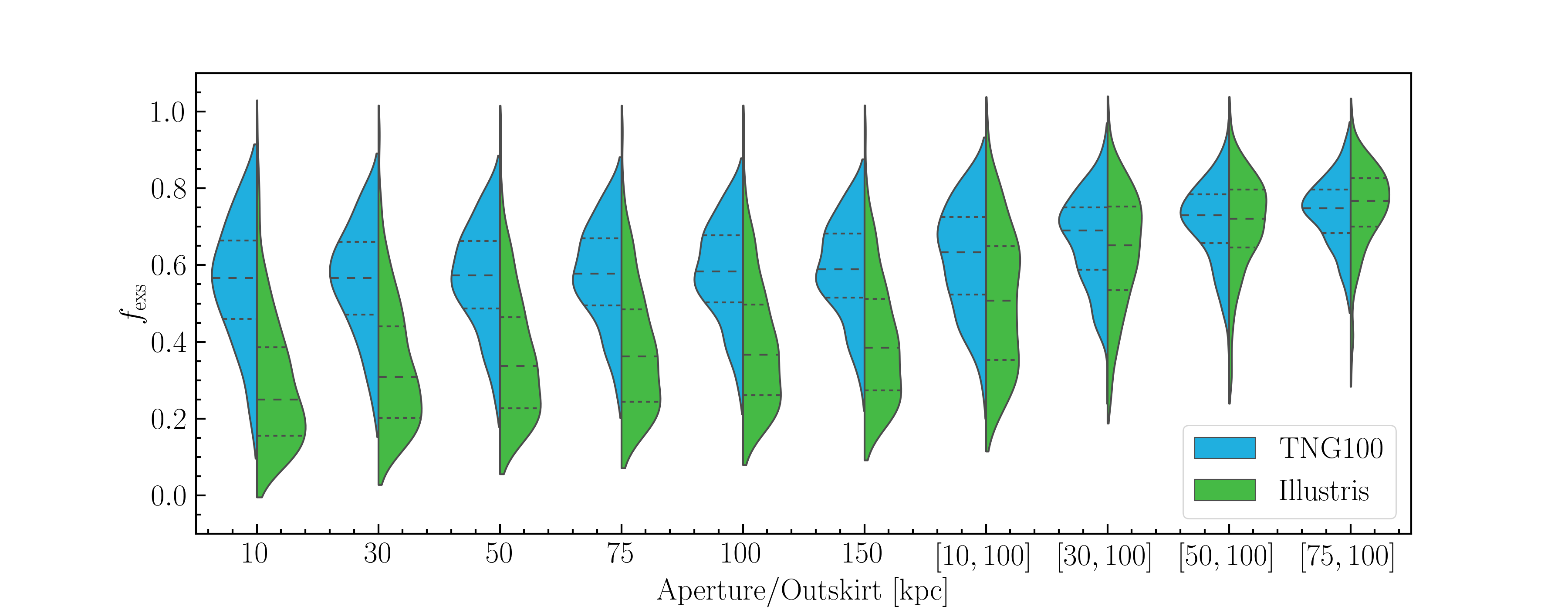}
        \caption{
            The distributions of \exs{} fraction ($f_{\rm exs}$) within different physical aperture or outskirt regions
            for galaxies in the {\tt TNG100} (blue, left) and {\tt Illustris} (green, right) simulations.
            We label the aperture size and the outskirt definitions in kpc on the X-axis. 
            We visualize the $f_{\rm exs}$ distributions using violin plots, and we also highlight their mean values (long-dashed line) and the 1-$\sigma$ ranges (dashed lines).
            Within different apertures, the $f_{\rm exs}$ distributions of {\tt TNG100} galaxies are clearly different from 
            the {\tt Illustris} ones with much higher mean $f_{\rm exs}$ values.
            This difference is most prominent in the inner 10 kpc. 
            However, the two simulations show remarkably similar $f_{\rm exs}$ distributions in the outskirt regions, 
            especially when the inner boundary is $\geq 30$ kpc.
            The \texttt{Jupyter} notebook for reproducing this figure can be found here: \href{https://github.com/Xuchuyi/HaloMassProxy/blob/main/FigureNotebook/Fig7.ipynb}{\faGithub}.
            }
       \label{fig:fig7}
    \end{figure*}

\subsection{Are our results robust against the choice of \changing{simulations}?}
    \label{ssec:diff}

    As with any work based on hydro-simulation, whether the results are robust against the choice of the simulations is always a valid question. While {\tt TNG}300 provides the volume and sample size of massive galaxies for the \topn{} tests, its lower mass resolution means that many low-mass satellites may become unresolved in the outskirts of massive galaxies or get disrupted earlier than in reality (e.g., \citealt{Springel2008MNRAS}; \citealt{Gao2012MNRAS}; \citealt{vandenBosch2018MNRAS}; \citealt{Bahe2019MNRAS}). In Fig.~\ref{fig:fig4}, the {\tt TNG}300 massive galaxies show systematically lower SHMR scatter, which could be caused by resolution. However, the SHMRs of the outskirt stellar mass with a $>50 $ kpc inner boundary are quantitatively consistent between {\tt TNG}300 and the other two higher-resolution simulations, illustrating the robustness of our results. 

    Meanwhile, it is well known that, compared to \illustris{}, the \tng{} series adopted many modifications and improvements (e.g., \citealt{Nelson2019MNRAS}) that help match the benchmark observations of massive galaxies (e.g., \citealt{Springel2018MNRAS, Nelson2018MNRAS, Genel2018MNRAS}). But the earlier onset and more powerful AGN feedback in \tng{} results in the earlier quenching of star-formation and more dominant \exs{} components in massive galaxies (e.g., \citealt{Tacchella2019MNRAS, Montenegro-Taborda2023, Cannarozzo2023MNRAS}). In Fig.~\ref{fig:fig7}, we directly compared the \exs{} stellar mass fractions within different apertures and outskirt regions explored in this work between \texttt{TNG}100 and \illustris{} massive galaxies. While the \exs{} fractions within all apertures confirm that the accreted stars dominated the entire massive galaxy in \texttt{TNG}100, even in the central 10 kpc, the \exs{} fractions in the outskirts are consistent between these two simulations. In either simulations, roughly 70\% of the \mout{50}{100} have an \exs{} origin. Although we only explored two hydro-simulations, the similar \exs{} fraction in the outskirt could help explain the robustness of our main conclusions against the detailed physical processes in hydro-simulations as it may suggest that the outskirt stellar mass is less sensitive to the physical recipes regulating the star formation and feedback processes but is more connected to the halo assembly itself. 
    
    Nevertheless, in Fig.~\ref{fig:fig4} and Fig.~\ref{fig:fig6} we can still notice subtle differences between \tng{}100 and \illustris{} simulations in the scatter values of the \mout{50}{100} SHMRs, which could provide some insight about massive galaxies' assembly (e.g., \citealt{Zentner2014MNRAS, Matthee2017MNRAS}). 
    \changing{With the help of new spectroscopic redshifts and weak lensing data for massive galaxies, we should soon be able to observationally constrain the scatter of SHMR at the high-mass end with much improved precision, which in return could serve as a diagnostic metric for validating hydro-simulations with different physical recipes for star formation and AGN feedback.}

    \begin{figure*}
        \centering
        \includegraphics[width=1.05\textwidth,trim=7cm 0 0 2cm]{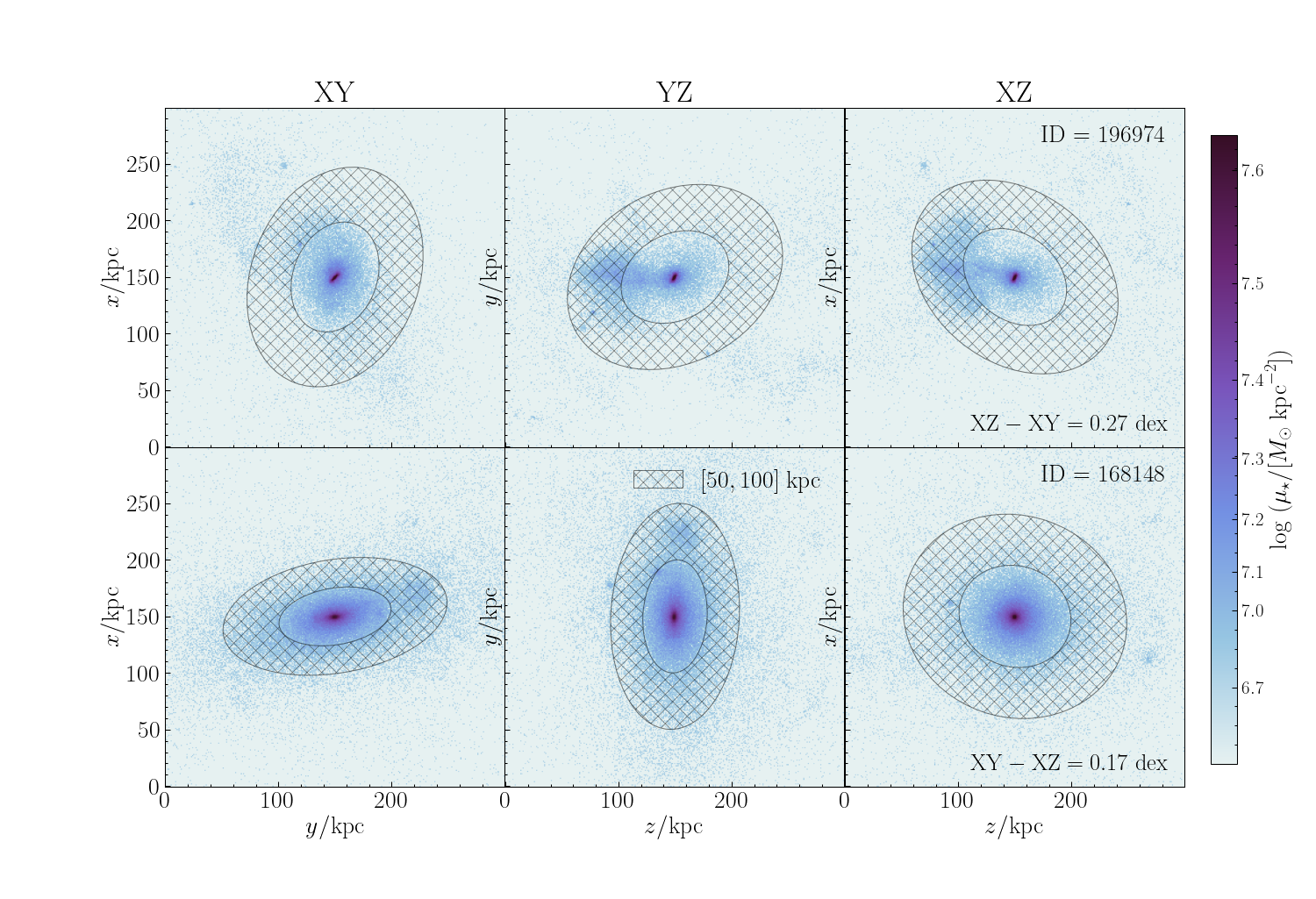}
        \caption{
            \change{Two example galaxies drawn from {\tt TNG}300 show prominent tidal features ({\bf top panels}) or prolate morphology ({\bf bottom panels}) impacting differently the outskirts in the three projections.}
            We show the 2-D stellar mass maps in three projections: {\tt XY} on the left, {\tt YZ} in the middle, and {\tt XZ} on the right. 
            On each map, we highlight the outskirt region used for its $M_{\star,\ [50,100]}$ measurement as in Figure \ref{fig:fig1} (hatched annulus).
            Using the measurement in the {\tt XZ} projection as a reference, we label the difference in $\log M_{\star,\ [50,100]}$ on the figure.
            {\bf Top panels} show galaxy ID$=196974$. As it is experiencing a major merger event, the asymmetric tidal features in the outskirts contribute to the strong dependence on projection.
            {\bf Bottom panels} show galaxy ID$=168148$ whose highly prolate intrinsic 3-D shape leads to strong projection dependence.
            The \texttt{Jupyter} notebook for reproducing this figure can be found here: \href{https://github.com/Xuchuyi/HaloMassProxy/blob/main/FigureNotebook/Fig8.ipynb}{\faGithub}.
            }
        \label{fig:fig8}
    \end{figure*}

\subsection{Is the outskirts stellar mass affected by projection effects?}
    \label{ssec:proj}

    One of the reasons to search for better halo mass proxies using the stellar mass distributions of the central galaxy is to eliminate or reduce the projection bias encountered by the richness-based method (e.g., \citealt{Sunayama2020MNRAS}; \citealt{Wu2022MNRAS}) when identifying galaxy clusters from imaging surveys as the property of a single galaxy will not be biased by the line-of-sight (LOS) large-scale structures or the photometric redshift uncertainties of nearby galaxies. However, massive galaxies are not spherical in 3-D (e.g., \citealt{Li2018ApJ}). Many massive galaxies' stellar halos show elongated (axis ratio $b/a<0.5$) shape in 2-D projection (e.g., \citealt{Huang2018}; \citealt{D'Souza2014MNRAS}) and could have prolate intrinsic 3-D shape (e.g., \citealt{Tsatsi2017A&A}; \citealt{Ene2018MNRAS}) induced partially by past mergers (e.g., \citealt{Li2018MNRASb}). While it is still challenging to observationally infer the distribution of massive galaxies' 3-D intrinsic shapes (but see, e.g., \citealt{Weijmans2014MNRAS}), it is safe to assume that the non-spherical 3-D shape will affect the physical meaning of the aperture or outskirt stellar masses estimated using 2-D images, which leaves room for ``single-galaxy projection bias''. 

    Statistically, massive galaxies' projections in the sky should be random over a large footprint (e.g., when the volume is much larger than a single galaxy cluster). In Fig.~\ref{fig:fig3}, we showed that the best-fit SHMR parameters in \texttt{TNG}300's three orthogonal projections are perfectly consistent, suggesting that the projection's impact on the mean SHMR is statistically isotropic, at least over the volume of the \texttt{TNG}300 simulation, which can be easily surpassed by modern cosmological imaging surveys. 
    However, this does not mean that the direction of our LOS to see the massive galaxy will not bias the stellar mass measurement.
    
    \change{In Fig.~\ref{fig:fig8}, we showcase two types of galaxies whose outskirt stellar mass (\mout{50}{100}) measurements have differences more significant than 0.15 dex among the three projections, which includes 581/2713 of the \texttt{TNG}300 sample.}
    The top panel shows an ongoing merger whose tidal features contribute significantly to the \mout{50}{100} when viewed from an advantageous projection ({\tt XY} and {\tt YZ} in this case). 
    Note that we create these stellar mass maps based on the FoF particles classified during the post-processing of the \change{hydrodynamical simulation}, which excludes particles belonging to satellites and stars in interacting halos. In practice, we will see the other \change{particles} of the merging system and need to mask or model them out before extracting the 1-D profile of the main galaxy to derive its \mout{50}{100}. This procedure could significantly reduce the impact of the projection angle. 

    Meanwhile, the bottom panel demonstrates the second category, whose stellar halo's intrinsic shape is highly prolate, and the 2-D axis ratio in the outskirt can be significantly different. In this case, the axis ratio drops from 0.89 in the more ``end-on'' {\tt XZ} projection to 0.50 (0.57) in the more ``edge-on'' {\tt XY} ({\tt YZ}) projections, while the $\log$\mout{50}{100} increases from 11.01 to 11.18 (11.18). In observation, our current procedure to estimate the outskirt stellar mass based on the average 1-D surface brightness profile \emph{along the major axis} (along an elliptical isophote that follows the 2-D shape) will be subject to this bias. Suppose such highly prolate cases constitute a significant \change{portion} of the massive galaxy population. In that case, our 2-D outskirt stellar mass measurements will ``favor'' a massive galaxy in a more ``edge-on'' LOS than those in a more ``end-on'' viewing angle, in the same sense as the red-sequence richness estimates ``favor'' the clusters with more nearby structures along the LOS \change{(\citealt{Sunayama2020MNRAS})}. Although the exact fraction of the highly prolate population among massive galaxies is still uncertain (e.g., \citealt{Li2018ApJ}; \citealt{Li2018MNRASb}; \citealt{Ene2018MNRAS}; \citealt{Thob2019MNRAS}), it warrants further investigation using the 3-D stellar mass distributions of massive galaxies in simulations, \change{which are presented in recent work (\citealt{Zhou2025})}. 

    In addition, by comparing the asymmetry and ellipticity of galaxies with \change{$\log$\mout{50}{100}} differences greater than 0.15 dex, we estimate the approximate fractions of the two types mentioned above. About $1/6$ of galaxies exhibit clear features of the first type (ongoing mergers), while around $4/5$ show characteristics of the second type (prolate shapes). The total fraction does not sum to unity because some galaxies display clear features of both types or neither.



\section{Summary and Conclusions} 
    \label{sec:summary}

    Recent observations of low-redshift massive galaxies with deep imaging data and high-quality lensing measurements point to an interesting connection between the outer stellar mass and their dark matter halos (e.g., \citealt{Huang2022}; \citealt{Montes2019MNRAS}).
    Hydro-simulations can generate statistically significant samples of massive galaxies at low redshift with reasonably realistic observational properties (e.g., \citealt{Pillepich2018MNRAS}; \citealt{Ardila2021MNRAS}). 
    This work uses the 2-D stellar mass distributions of massive central galaxies from the \illustris{} and \tng{} simulations to confirm the connection between outskirt and halo mass. We further discuss potential applications and investigate the physical origin of this relation. 
    We evaluated the performance of different measurements of aperture and outskirt stellar masses as halo mass ($M_{200{\rm c}}$) proxies using 1) the slope and scatter of their best-fit stellar-to-halo mass relations, 2) the scatter values of halo mass ($\sigma_{\log M_{\rm halo}}$) of samples selected within the same ranges of cumulative number density ranges, also known as the \topn{} tests defined in \citet{Huang2022}.
    We also consider these massive galaxies' \ins{} and \exs{} components. 
    Inspired by the popular ``two-phase'' formation scenario of massive galaxies and recent results based on semi-empirical models (e.g., {\tt UniverseMachine}, \citealt{Bradshaw2020MNRAS}), we compared the performance of \exs{} stellar mass as a halo mass proxy to different aperture or outskirt stellar masses.
    We also examined the \exs{} fraction in a series of apertures and within different outskirt regions.
    Our main findings are:
    
    \begin{itemize}
    
        \item Using low-redshift ($z\sim 0.4$) massive galaxies in all three hydro-simulations, \illustris{}, {\tt TNG}100 and {\tt TNG}300, we confirm that their outskirt stellar masses (e.g., stellar mass between 50 to 100 kpc using an elliptical aperture)  correlates better with halo mass than all aperture stellar masses. This conclusion stands even for the very large aperture stellar mass (e.g., within a 150 kpc aperture). This result is driven by the slope of the SHMR for outskirt stellar masses, which is significantly shallower ($\sim 1$) than other aperture stellar masses' ($\sim 1.5$)\footnote{When using the observable (stellar mass) as the X-axis}. This conclusion is robust against the different physical recipes for galaxy formation and different mass resolutions among the three simulations examined here. This suggests that the extended stellar halos of low-$z$ massive galaxies preserve precious information about dark matter halo assembly and can serve as a more promising halo mass proxy than the ``total'' stellar mass (Fig.~\ref{fig:fig4})
        
        \item Following the strategy outlined in \citet{Huang2022}, we conduct \topn{} tests by rank ordering a large sample of massive galaxies from the {\tt TNG}300 simulation based on different measurements of aperture and outskirt stellar masses. We then select these galaxies into different \topn{} bins using consistent thresholds or specific rank ranges, such as the top 100 or the top 50 to 100 most massive galaxies based on a particular stellar mass metric. Then, we compare the scatter of halo mass across the different samples. The results (Fig.~\ref{fig:fig5}) are qualitatively consistent with the $0.2 < z < 0.5$ HSC massive galaxies, which confirms that the \topn{} strategy is a straightforward and practical approach to inter-compare various tracers as halo mass proxies. Meanwhile, the scatter values from {\tt TNG}300 are systematically lower than those inferred in observation using galaxy-galaxy lensing data and N-body simulations. And, the \topn{} tests in this work using consecutive cumulative number density (ranks) bins do not reveal the same increasing trend between halo mass scatter and ranking thresholds as in observation.
        
        \item In the {\tt TNG} simulations, we find that the outskirt stellar mass (e.g., \mout{50}{100}) can even outperform the total \exs{} stellar mass as a halo mass proxy (Fig.~\ref{fig:fig6}). This seemingly counter-intuitive result indicates an interesting ``decoupling'' between the stellar and halo masses' assembly, where a considerable fraction of the \emph{technically} \exs{} stellar content from early major mergers may not show a robust physical connection with today's halo mass. On the other hand, the outskirt stellar mass could represent the stars accumulated from minor mergers for an extended period and correlate well with the `historic richness', or the cumulative abundance of sub-halos in a significant fraction of the assembly history of a massive dark matter halo.
        
        \item We also show that, while the dramatically different baryonic physical recipes adopted in \illustris{} and {\tt TNG}100 lead to significantly different \exs{} fractions within the inner regions of massive galaxies (e.g., within 30 kpc), the \exs{} fractions within the outskirt regions (e.g., between 50 and 100 kpc) are statistically consistent with each other (Fig.~\ref{fig:fig7}). 
        This result may suggest that, as a halo mass proxy, the outskirts stellar mass could be less sensitive to the complex physical processes shaping the central region, such as the feedback from a supermassive black hole.
        
    \end{itemize}

    These promising results motivate us to pursue a series of near-future directions to advance the exploration of the physical connection between different parts of massive galaxies and their halo assembly history, to establish a more comprehensive galaxy-halo connection model for galaxy formation and cosmology.
    \changing{With the arrival of deep imaging data from Euclid and the Vera C. Rubin Observatory's Legacy Survey of Space and Time (LSST), we will be able to characterize the ubiquitous diffuse stellar component in massive dark matter halos with unprecedented precision and detail, presenting an excellent opportunity to investigate the scientific potential of the ICL or the extended stellar halos around massive galaxies.}
    
    Using hydro-simulations, we will further study galaxy outskirts, focusing on mass resolution effects (Leidig et al. in preparation). We will also compare the observed aperture or outskirt stellar masses with those defined in 3D to understand the systematics induced by the single-galaxy projection effect and the intrinsic shape of stellar halos (\citealt{Zhou2025}). 
    More importantly, as the richness of red-sequence (quenched) galaxies in the dark matter halo is still the most commonly adopted halo mass proxy in the galaxy cluster community, we will compare different definitions of richness' performance against the outskirt stellar mass' based on \change{hydrodynamical simulations} and investigate the physical differences between the current and `historic' richness values (Xu et al. in preparation). 
    Furthermore, instead of relying on a single outskirts stellar mass value, we will use an empirical model to connect the halo assembly history to the gradual accumulation of the extended stellar halo and the complete stellar mass profiles of massive galaxies. 
    With the help of a much larger sample of massive galaxies, deeper \& better images, higher-quality weak lensing data \change{(e.g., from HSC, LegacySurvey, Euclid, and Vera C.Rubin Observatory)}, and more precise redshift measurements (e.g., from DESI), we will attempt to apply such an empirical model to gain a more comprehensive understanding of galaxy-halo connection at the most massive end.  



\section*{Acknowledgment}

  SH acknowledges support from the National Natural Science Foundation of China Grant No. 12273015 \& No. 12433003 and the China Crewed Space Program through its Space Application System. 
  
  This material is based upon work supported by the National Science Foundation under
  Grant No. 1714610.
  This research was supported in part by the National Science Foundation under Grant number 2206695. BD gratefully acknowledges support by the Sloan Foundation.

  The authors acknowledge support from the Kavli Institute for Theoretical Physics. The National Science Foundation under Grant No. NSF PHY11-25915 also partly supported this research and Grant No. NSF PHY17-48958

  We acknowledge the use of the lux supercomputer at UC Santa Cruz, funded by NSF MRI grant AST
  1828315. AL is supported by the U.D Department of Energy, Office of Science, Office of High
  Energy Physics under Award Number DE-SC0019301. AL acknowledges support from the David and
  Lucille Packard Foundation, and from the Alfred. P Sloan Foundation.
  
\software{
    \href{http://www.numpy.org}{\code{NumPy}} \citep{Numpy},
    \href{https://www.astropy.org/}{\code{Astropy}} \citep{astropy},  
    \href{https://www.scipy.org}{\code{SciPy}} \citep{scipy}, 
    \href{https://matplotlib.org}{\code{Matplotlib}} \citep{matplotlib} 
}



\bibliography{exsitu}{}
\bibliographystyle{aasjournal}

\end{CJK*}
\end{document}